\documentclass[aps,pre,preprint,superscriptaddress,preprintnumbers,showpacs,nofootinbib]{revtex4-1}
\usepackage{amsmath,amssymb}
\usepackage[dvipdf,dvips,dvipdfmx]{graphicx}
\usepackage{here}
\usepackage{color}
\usepackage{url}
\usepackage{algorithm}
\usepackage{algorithmic}
\usepackage[subrefformat=parens]{subcaption}
\newcommand{\cc}{\mathrm{c}}
\newcommand{\M}{\mathcal{M}}

\newcommand{\Tr}{\mathrm{Tr}}

\makeatletter

\begin{document}

\title{Triad second renormalization group}

\author{Daisuke Kadoh}
\affiliation{
Faculty of Sciences and Engineering, Doshisha University, Kyotanabe, Kyoto 610-0394, Japan}
\affiliation{Research and Educational Center for Natural Sciences, Keio University, Yokohama 223-8521, Japan}

\author{Hideaki Oba}
\email[]{h\_oba@hep.s.kanazawa-u.ac.jp}
\affiliation{Institute for Theoretical Physics, Kanazawa University,
 Kanazawa 920-1192, Japan}

\author{Shinji Takeda}
\affiliation{Institute for Theoretical Physics, Kanazawa University,
 Kanazawa 920-1192, Japan}

\date{\today}

\begin{abstract}
  We propose a second renormalization group (SRG) in the
 triad representation of tensor networks.  
The SRG method  improves two parts of the triad tensor renormalization group, which are the decomposition 
of intermediate tensors  and the preparation of isometries,  taking the influence of environment tensors into account.
Every fundamental tensor including environment tensor is given as
a rank-3 tensor, and the computational cost of the proposed algorithm scales with ${\cal O}(\chi^5)$ employing the randomized SVD
where $\chi$ is the bond dimension of tensors. 
  We test this method in the classical Ising model on the two dimensional square lattice,
and find that numerical results are obtained in good accuracy for a fixed computational time.
\end{abstract}

\preprint{KANAZAWA-21-03}

%title
\maketitle

\section{Introduction\label{sec:Introduction}}
The tensor network method is a promising approach in investigating 
quantum and classical many-body systems\cite{
White:1992zz,
DMRGreview,MPS,PEPS,TNrepresentation,Liu:2013nsa,
Banuls:2016gid}.
The tensor renormalization group (TRG) \cite{Levin:2006jai} 
provides accurate results in practice for two dimensional classical models 
from condensed matter physics to lattice field theory \cite{Shimizu:2012zza,Yu:2013sbi,Shimizu:2014uva,Kawauchi:2016xng,Kadoh:2018tis}, 
even for theories with the sign problem \cite{Denbleyker:2013bea,Shimizu:2014fsa,Shimizu:2017onf,Kadoh:2018hqq,Kadoh:2019ube}. 
In two dimensions, 
there are also powerful methods with disentangler\cite{TNR,loopTNR,rec_trg_1,rec_trg_2}
by which the accuracy of results are significantly improved even at criticality.
In dimensions higher than two,  several algorithms such as the higher-order TRG (HOTRG) \cite{HOTRG}, anisotropic TRG (ATRG) \cite{Adachi:2019paf}
and the triad TRG \cite{Kadoh:2019kqk}
were proposed and used in recent works \cite{Akiyama:2020ntf,Akiyama:2020soe,Akiyama:2021zhf}. 
Further improvements of the algorithms would be necessary in obtaining accurate results 
because the truncation error becomes larger in higher dimensions
as well as  
the computational cost does.

In the TRG, the tensor  is  decomposed locally by the singular value decomposition (SVD), which 
does not always give the best approximation of the partition function itself
because the influence of the rest of networks (i.e.\ environment tensors) is ignored. 
Isometries are also optimized locally by using the SVD in some algorithms.
For higher dimensional theories,  
the influence of environment tensors  should 
be treated much more carefully 
because the size of environment lattice increases with volume $V=L^d$ for fixed $L$.

The second renormalization group (SRG) method \cite{Xie:2009zzd} 
incorporates the influence of environments into the decompositions of tensors
so that the partition function 
 is well approximated, and the truncation error is drastically reduced.  
 Similar improvements have been done in the HOSRG \cite{HOTRG}.
Although the SRG-scheme is very powerful,  
its computational cost is higher than that of the local algorithms. 
So it is important to realize the SRG in efficient algorithms.

The triad TRG method \cite{Kadoh:2019kqk} is formulated 
on a tensor network made only of rank-3 tensors, 
which is referred to a triad network in this paper.  
The HOTRG-like renormalization is easily 
transcribed on the triad network, 
and the computational cost is reduced in higher dimensions. 
If the environment tensors are also given by the similar triad representation, 
the SRG-scheme could work within  
a reasonable cost in higher dimensions. 
However, it is still an open question whether the concept of SRG and 
rank-3 environment tensors coexist within good accuracy even in two dimensions.

This paper is devoted to address this issue.
We present the SRG method on a two dimensional triad network. 
All fundamental tensors including environment tensors are given as
rank-3 tensors. 
The SRG scheme is used to improve two parts of the triad TRG, 
which are the decomposition of intermediate tensors and making isometries. 
The forward and backward schemes adopted  in the HOSRG 
are given with the rank-3 environment tensors.
The computational cost scales with ${\cal O}(\chi^6)$ where $\chi$ is the bond dimension of the tensor,
 and it is reduced to ${\cal O}(\chi^5)$ by using the randomized SVD. 
We find that our method named a triad SRG effectively works within good accuracy.

The rest of this paper is organized as follows.
In Sec.~\ref{sec:SRG}, 
we begin with reviewing the TRG/SRG on a honeycomb lattice, and 
see the HOTRG/HOSRG schemes defined on a square lattice in detail. 
In Sec.~\ref{sec:Triad_SRG}, 
the triad SRG methods are presented. Reviewing the triad TRG, an algorithm of the triad SRG whose cost is $\chi^6$ is firstly given. 
Then a $\chi^5$-algorithm is derived from the $\chi^6$ one by using the randomized SVD. 
In Sec.~\ref{sec:numerical_results},
we show numerical results of 
the square lattice Ising model and discuss the performance
by comparing the triad SRG to the other methods. 
A summary and a discussion for the extension to a higher dimensional system
are given in Sec.~\ref{sec:discussion}.

Our main purpose in this paper is to pursue how the idea of small connectivity is compatible with a philosophy of SRG
looking ahead to future applications to higher dimensional systems.

\section{Second renormalization group method}
\label{sec:SRG}

\subsection{TRG/SRG on a honeycomb lattice\label{sec:TRGandSRG}}

\begin{figure}[H]
\centering
\includegraphics[width=160mm]{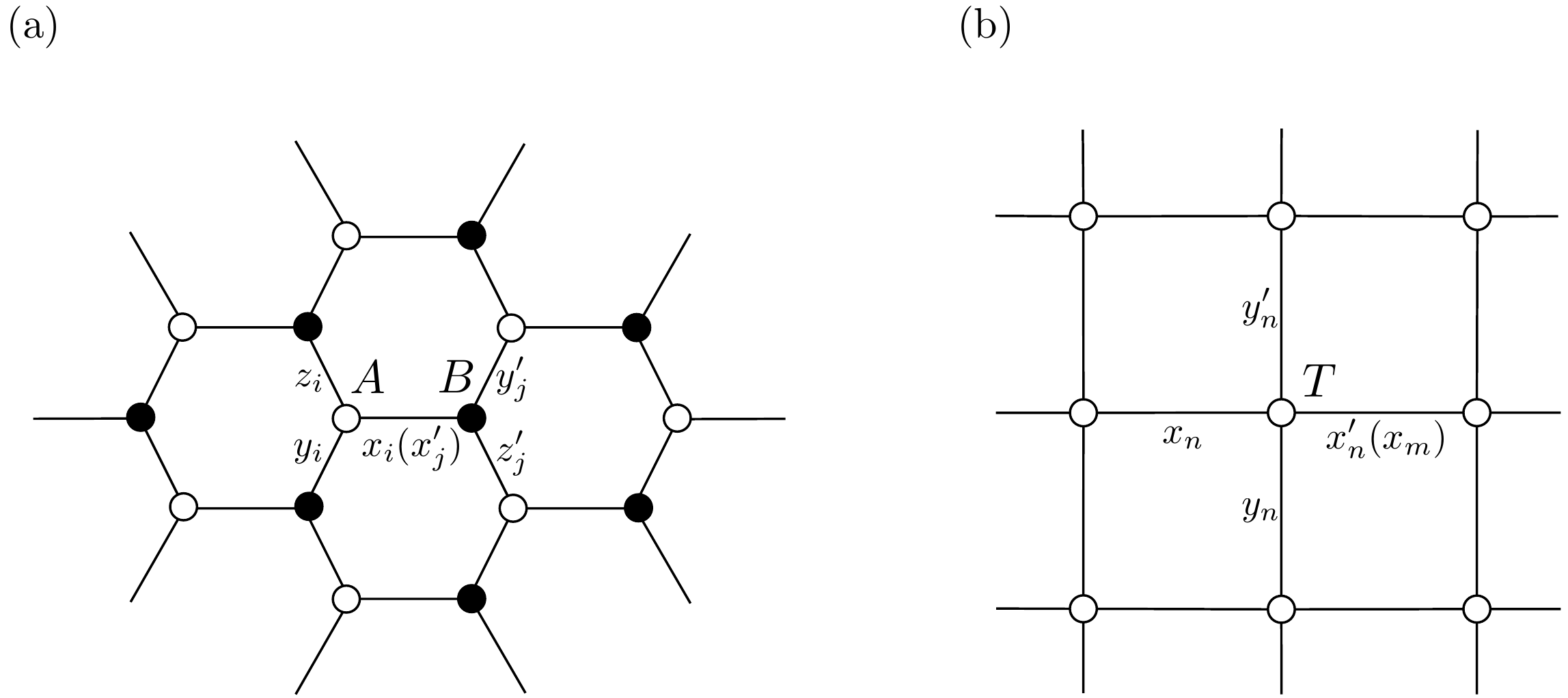}
\caption{
Tensor network on a honeycomb lattice (a) and 2d square lattice (b). 
}
 \label{fig:honeycomb}
\end{figure}

We begin with considering 
a homogeneous tensor network on a honeycomb lattice: 
\begin{align}
  Z = \Tr \prod_{i \in w, j\in b} A_{x_i y_i z_i} B_{x'_j y'_j z'_j},
  \label{TN_honeycomb}
\end{align}
where $b$ and $w$ are the sets of black and white lattice points, and 
$\Tr$ is the summation over all tensor indices $x,y, z=1,2,\cdots,\chi$. 
The link $x_i$ is identified to $x'_j$ where $j$ is the nearest neighbor site of $i$ in the $x$ direction,
and the same applies to $y'$ and $z'$.
Fig.~\ref{fig:honeycomb} (a) shows the tensor network on the honeycomb lattice. 
We assume that $A$ and $B$ are totally symmetric tensors for simplicity of explanation.
It is straightforward to extend the algorithm presented in this section to the non-symmetric case.

The singular value decomposition (SVD) of an $n \times n$ matrix $M$, 
which is used for the coarse graining, 
 is defined  in the standard manner as  
\begin{align}
  M = U \Lambda V^\dag, 
  \label{SVD}
\end{align}
where $\Lambda$ is an $n \times n$ diagonal matrix in which 
singular values are sorted in the descending order,  and $U$ and $V$ are $n \times n$ unitary matrices. 
Taking the $k$ largest singular values ($k<n$) gives 
the best $k$-rank approximation of $M$ as
\begin{align}
  M_{IJ} \approx \sum_{m=1}^k C_{Im} D_{Jm} 
  \label{low_rank_approaximation}
\end{align}
where $C_{Im} = (U\sqrt{\Lambda})_{Im}$ and $D_{Jm} = (\sqrt{\Lambda} V^\dag)_{mJ}$.

\begin{figure}[H]
\centering
\includegraphics[width=80mm]{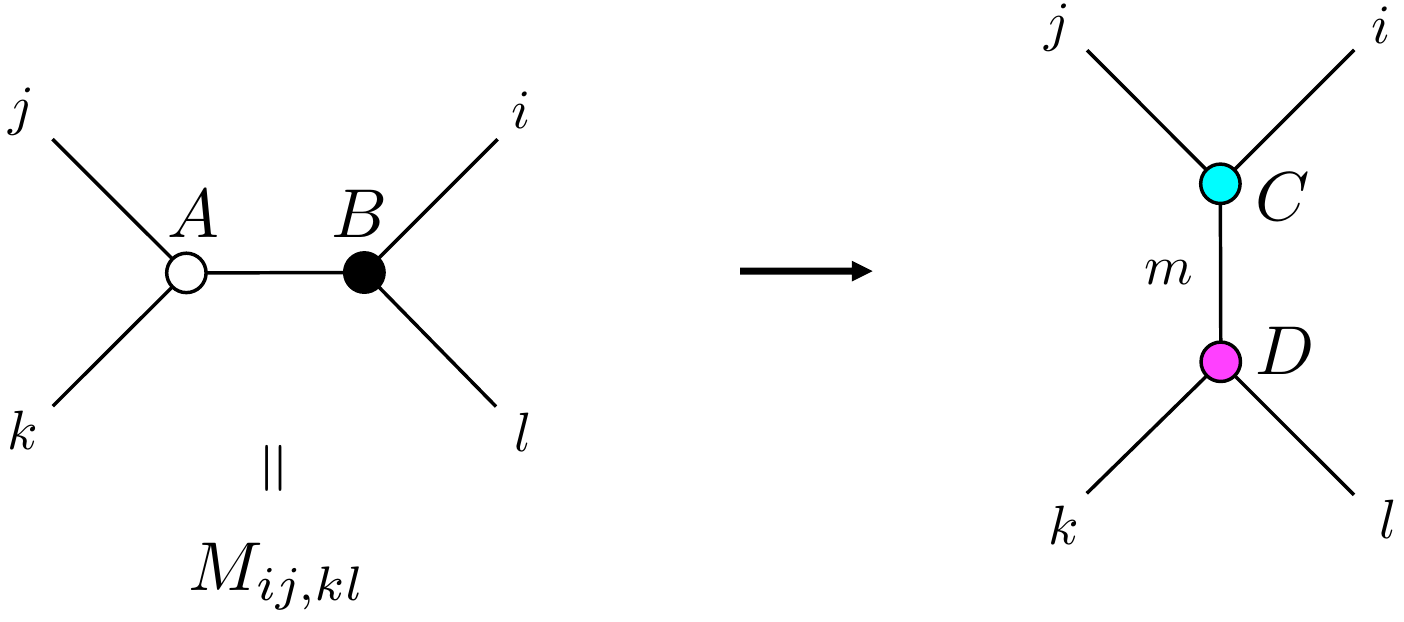}
\caption{Transformation of $M$ by SVD. }
 \label{fig:svd_M}
\end{figure}

\begin{figure}[H]
\centering
\includegraphics[width=140mm]{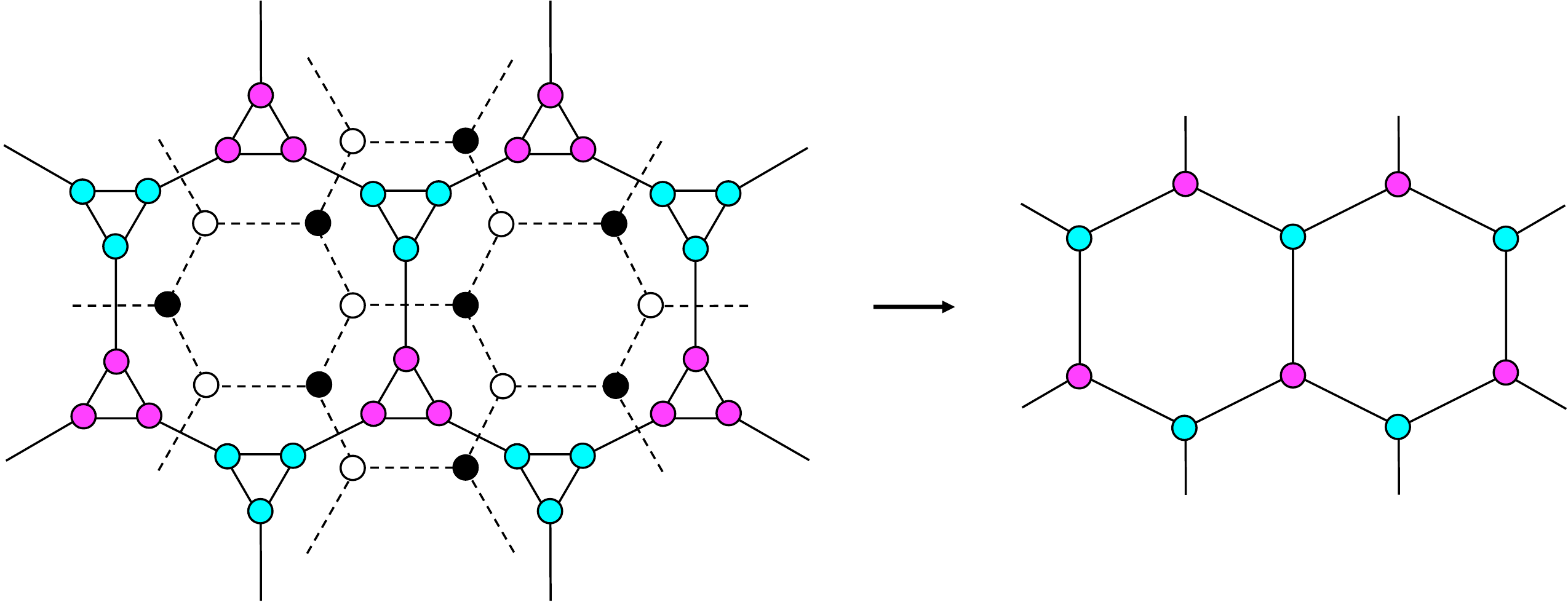}
\caption{
TRG on the honeycomb tensor network. 
}
 \label{fig:renomalization_honeycomb}
\end{figure}

The partition function is evaluated by the TRG using the SVD such that the form of Eq.~(\ref{TN_honeycomb}) is 
kept at $n$th renormalization step with $A^{(n)}$ and $B^{(n)}$.
Consider a $\chi^2 \times \chi^2$ matrix $M^{(n)}$ defined by
\footnote{
The comma in $M_{ij,kl}$ is employed to regard a tensor $M_{ijkl}$ as a matrix with the column $i,j$ and the row $k,l$.  
Unless otherwise noted, $\sum_i$ denotes $\sum_{i=1}^\chi$ in this paper.
}
\begin{align}
  M^{(n)}_{ij,kl} \equiv \sum_m A^{(n)}_{mkj} B^{(n)}_{mil}. 
  \label{M_honeycomb}
\end{align}
With the Eqs.~(\ref{SVD}) and (\ref{low_rank_approaximation}), we have a $\chi$-rank approximation of $M^{(n)}$ as
\begin{align}
M^{(n)}_{ij,kl} \approx \sum_{m=1}^\chi C^{(n)}_{ijm} D^{(n)}_{klm}. 
\label{M_local_decomposition}
\end{align}
Fig.~\ref{fig:svd_M} shows the transformation of $M^{(n)}$
from Eq.~(\ref{M_honeycomb}) to (\ref{M_local_decomposition}).
As shown in Fig.~\ref{fig:renomalization_honeycomb},
we arrive at Eq.~(\ref{TN_honeycomb}) again with
renormalized tensors,
\begin{align}
& A^{(n+1)}_{xyz} \equiv \sum_{i,j,k} C^{(n)}_{ijx}  C^{(n)}_{jky}  C^{(n)}_{kiz}, \nonumber \\
& B^{(n+1)}_{xyz} \equiv \sum_{i,j,k} D^{(n)}_{ijx} D^{(n)}_{jky} D^{(n)}_{kiz}.
\label{renormlized_AB}
\end{align}
In case of finite volume lattice of $V=2 \times 3^N$ with the center symmetric boundary condition, 
the renormalization is completed in $N$ steps. 
Finally we can evaluate $Z$ from two tensors $A^{(N)}$ and $B^{(N)}$ as 
 $Z \approx \sum_{x,y,z} A^{(N)}_{xyz}B^{(N)}_{xyz}$
where $A^{(0)}$ and $B^{(0)}$ are the initial tensors. 

Although the SVD gives the best approximation of tensors locally, 
it does not always give the best one for $Z$
because the rest of tenor network (i.e. the environment for $M$) is ignored. 
The second renormalization group (SRG) method incorporates the influence of environments 
into the decomposition of $M$ 
so that $Z$ is well approximated. 

\begin{figure}[H]
\centering
\includegraphics[width=80mm]{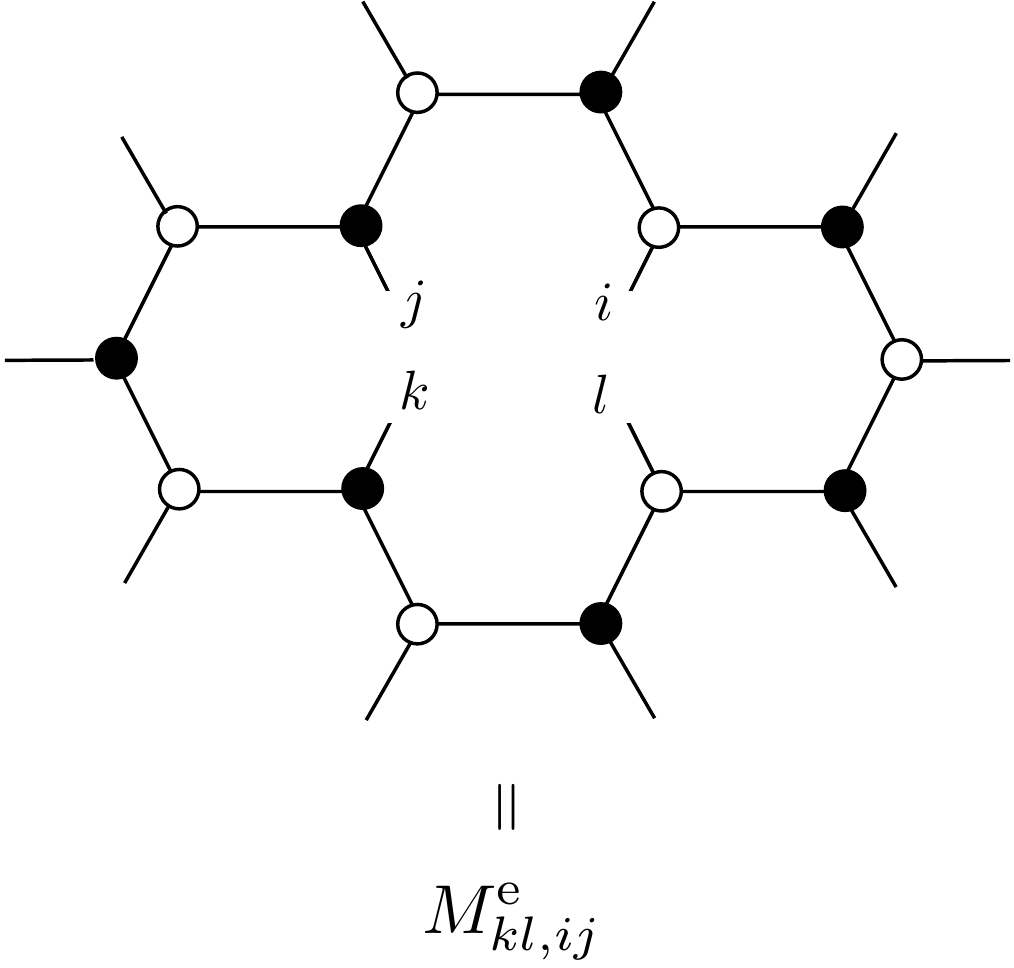}
\caption{The environment of $M$ in Eq.~(\ref{M_honeycomb})
($M^{\rm e}$).
}
 \label{fig:environment_tensor_honeycomb}
\end{figure}

In order to formulate the SRG, let us express $Z$  as 
\begin{align}
Z = {\rm Tr}(MM^{\rm e}) \equiv \sum_{i,j,k,l} M_{ij,kl} M^{\rm e}_{kl,ij}
\end{align}
where $M$ is given as Eq.~(\ref{M_honeycomb}) from $A$ and $B$.  
As shown in Fig.~\ref{fig:environment_tensor_honeycomb},
$M^{\rm e}$ is the rest tensor network excluding $M$ from $Z$. 
It is difficult to evaluate $M^{\rm e}$ exactly for a large volume lattice 
because $M^{\rm e}$ contains many tensor contractions.

\begin{figure}[H]
\centering
\includegraphics[width=160mm]{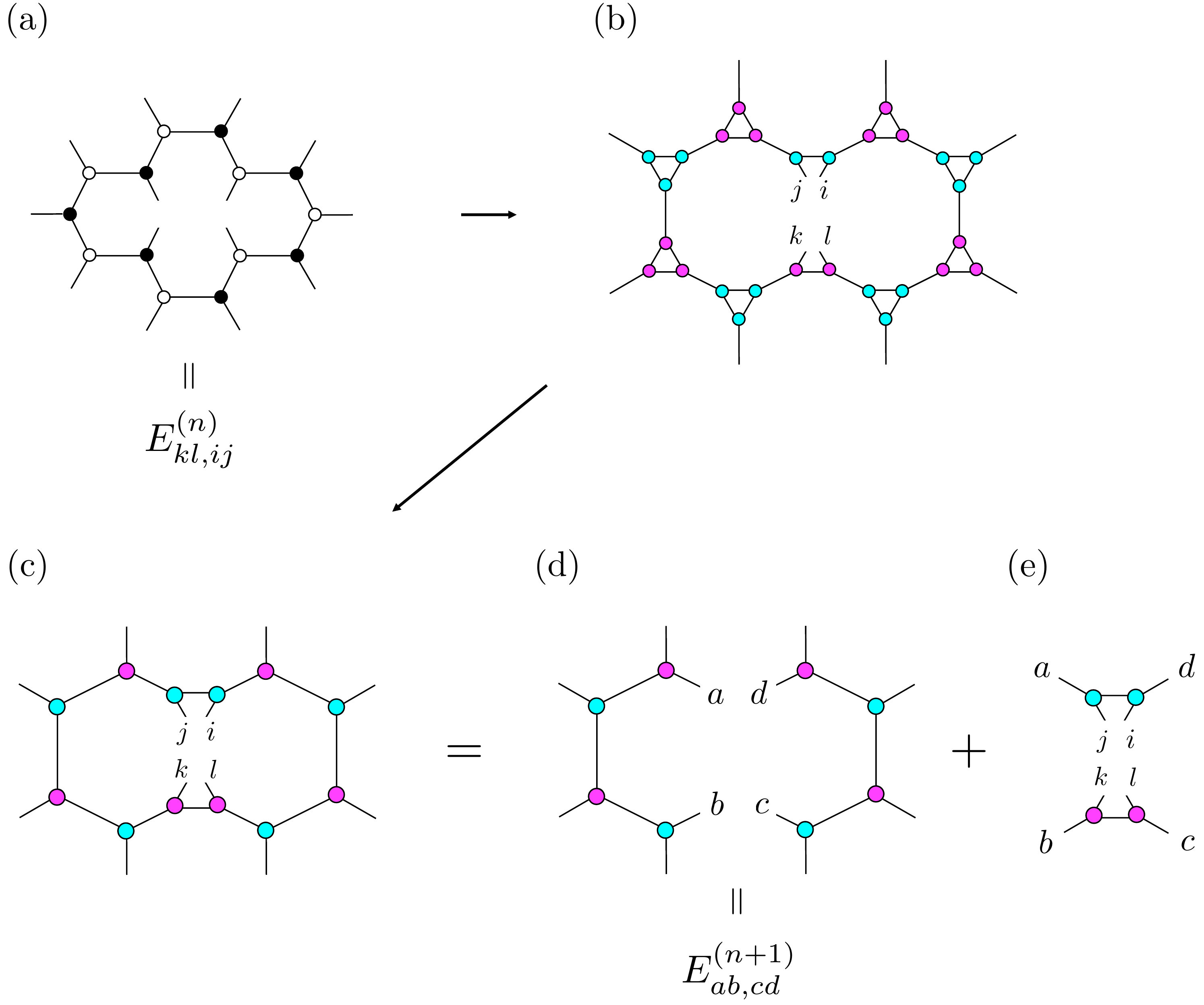}
\caption{
Renormalization of environment tensors in SRG. 
}
 \label{fig:renormalization_environments}
\end{figure}

\begin{figure}[H]
\centering
\includegraphics[width=60mm]{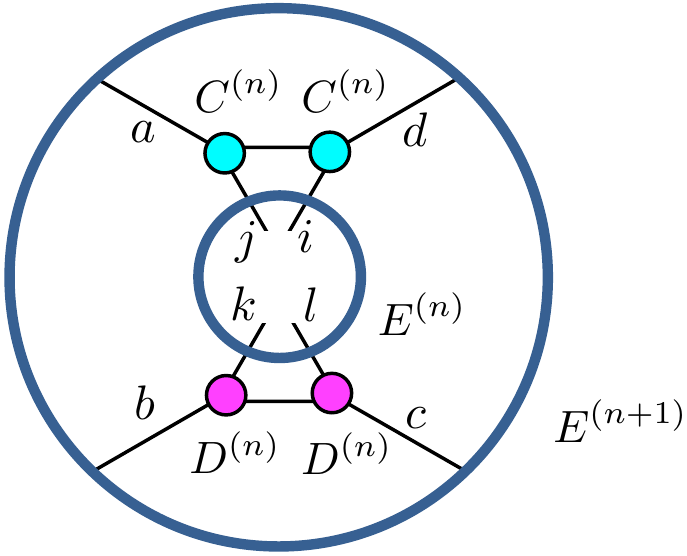}
\caption{
Schematic representation of the recurrence relation Eq.~(\ref{recurrence_relation_srg}). 
}
 \label{fig:reurrence_relation}
\end{figure}

One may evaluate $M^{\rm e}$ iteratively using the TRG method as shown in Fig.~\ref{fig:renormalization_environments} 
where $E^{(0)}=M^{{\rm e}}$ at the first renormalization step.
The coarse-grained environments approximately satisfy a recurrence relation, 
\begin{align}
E^{(n)}_{kl,ij}=\sum_{a,b,c,d,e,f} E^{(n+1)}_{ab,cd} C^{(n)}_{jea} C^{(n)}_{eid} D^{(n)}_{kfb} D^{(n)}_{flc}.
\label{recurrence_relation_srg}
\end{align}
We rather define the $i$-th scale environment tensor $E^{(i)}$ with Eq.~(\ref{recurrence_relation_srg})  
by solving it $N-1$ times from $E^{(N)}_{ab,cd} = \delta_{ac} \delta_{bd}$ (backward step)
and finally obtain $E^{(0)}$ which is an approximate estimate of $M^{\rm e}$.
Before starting the backward step, we perform the usual TRG procedure (the forward step)
to generate $C^{(n)}$ and $D^{(n)}$ ($n=1,2,...,N-1$).
See Fig.~\ref{fig:reurrence_relation} for a schematic representation of the recurrence relation of Eq.~(\ref{recurrence_relation_srg}).
In the figure we have to take care that the bold circle with internal four links are not four rank-3 tensors 
but a rank-$4$ environment tensor $E^{(n+1)}$.
We should also note that  $E^{(n+1)}$ has 
coarse grained indices which emerge when dropping
Fig.~\ref{fig:renormalization_environments} (e). 

Once $M^{\rm e}$ is obtained, $M$ 
is decomposed by the fundamental procedure of SRG as follows.  
We firstly decompose $M^{\rm e}$ by the SVD as $M^{\rm e} =  U_{\rm e} \Lambda_{\rm e} V_{\rm e}^\dag$
and define 
\begin{align}
\tilde M \equiv \sqrt{\Lambda_{\rm e}} V_{\rm e}^\dag M U_{\rm e} \sqrt{\Lambda_{\rm e}}. 
\label{def:tildeM}
\end{align}
Then we also decompose $\tilde M$ by the SVD as
\begin{align}
\tilde M = \tilde U \tilde \Lambda \tilde V^\dag.
\label{svd_tildeM}
\end{align}
Since $Z={\rm Tr}\, \tilde M$, the (truncated) SVD of $\tilde M$ gives the best approximation of $Z$.  
From Eq.~(\ref{def:tildeM}) and Eq.~(\ref{svd_tildeM}), we have an approximation of $M$ as
\begin{align}
M_{IJ} \approx \sum_{m=1}^\chi C_{Im} D_{Jm},  
\label{good_svd_M}
\end{align}
where $C_{Im}= (V_{\rm e} \sqrt{\Lambda_{\rm e}}^{-1} \tilde U \sqrt{\tilde \Lambda})_{Im}$
and  $D_{Jm} =  ( \sqrt{\tilde \Lambda} \tilde V^\dag  \sqrt{\Lambda_{\rm e}}^{-1} U_{\rm e}^\dag )_{mJ}$. 
Applying these procedures in Eq.~(\ref{def:tildeM})--(\ref{good_svd_M})
to $M^{(n)}$ given by Eq.~(\ref{M_honeycomb}),
the truncation error of the whole tensor network is minimized,
rather than that of $M^{(n)}$.

In this way, the SRG for the honeycomb lattice \cite{Xie:2009zzd}
can capture the environment effects and reduce the truncation error in the coarse-graining steps.
The computational costs of TRG and SRG scale with $\chi^6$ 
since the SVD and the recurrence relation have a scaling of $\chi^6$   
where the cost of SVD  is reduced to $\chi^5$ by using the randomized SVD, and the cost of TRG can be reduced in this sense. 
Numerical results presented in Fig.~4 of  Ref. \cite{Xie:2009zzd} show 
that the error of TRG is drastically reduced by the SRG.

\subsection{HOTRG/HOSRG on a 2d square lattice\label{sec:HOTRGandHOSRG}}

We consider a homogeneous tensor network of rank-4 tensors $T_{ijkl}$ ($i,j,k,l=1,\ldots, \chi$) 
on the two dimensional square lattice : 
\begin{align}
  Z = \Tr \prod_{n \in \Gamma} T_{x_n x'_n y_n y'_n}.
  \label{tensor_network}
\end{align}
where $\Gamma$ is the set of all lattice sites.
As can be seen in Fig.~\ref{fig:honeycomb} (b), 
$x'_n$ coincides with the index $x_m$ of the tensor on the nearest neighbor site $m$ in the positive
$x$ direction, and the same applies to $y'_n$.  
Therefore each summation is the contraction over the shared indices. 
  
\begin{figure}[H]
\centering
\includegraphics[width=150mm]{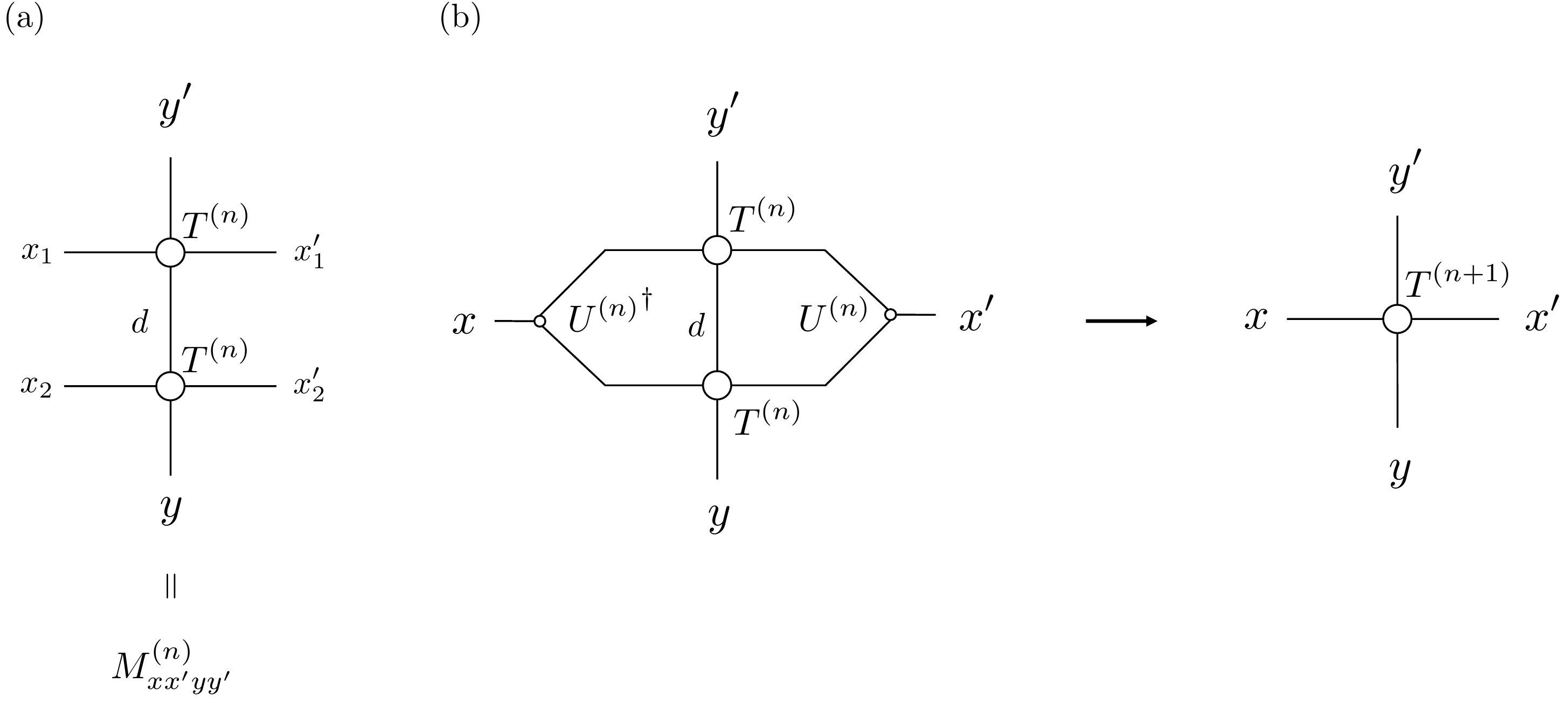}
\caption{
(a) Graphical representation of $M$ in the HOTRG.
(b) Renormalization of the HOTRG. 
}
 \label{fig:M_of_HOTRG}
\end{figure}

The renormalization of HOTRG is carried out for $x$ and $y$ directions alternately. 
We first consider the renormalization along the $y$ axis.  
Let $M^{(n)}$ be rank-6 tensor made of two tensors $T^{(n)}$ as
\begin{align}
M^{(n)}_{xx'yy'} \equiv \sum_d T^{(n)}_{x_1x_1'dy'} T^{(n)}_{x_2x_2'yd} 
\label{M_HOTRG}
\end{align}
where $x=x_1\otimes x_2$ and  $x'=x_1'\otimes x_2'$, which is shown 
in Fig.~\ref{fig:M_of_HOTRG} (a). 

Let us consider a matrix 
representation of $M$ as  $M'_{x,x'yy'} \equiv M_{xx'yy'}^{(n)}$ to create an isometry $U$ 
by which the tensor network is renormalized.  We diagonalize $K \equiv M' M'^{\dag}$ as
\begin{align}
K = U^{(n)} \Lambda^{(n)} {U^{(n)}}^\dag
\label{left_unitary}
\end{align}
where ${\Lambda}$ is a diagonal matrix in which eigenvalues $\lambda_i$ are
sorted in the descending order, and $U$ is the unitary matrix. 
We employ $U$ as the isometry
\footnote{
We actually choose the isometry $U$ from two unitary matrices: 
the left unitary matrix 
$U_L$ defined by Eq.~(\ref{left_unitary})
and the right unitary matrix $U_R$ defined by diagonalizing $\tilde K=M'{M'}^\dag$ for $M'_{x',xyy'} \equiv M_{xx'yy'}$
as $\tilde K= U_R \Lambda_R {U_R}^\dag$. 
Then $U$ is chosen to be $U=U_L$ for $\epsilon_L \le \epsilon_R$ 
and  $U=U_R^\ast$ for $\epsilon_R < \epsilon_L$ where $\epsilon_Q \equiv \sum_{i>\chi} \lambda_{Q,i}$ for $Q=L,R$.  
See Ref. \cite{HOTRG} for the detail. 
}
and define a renormalized tensor $T^{(n+1)}$ as
\begin{align}
T^{(n+1)}_{yy'xx'} \equiv \sum_{i,j=1}^{\chi^2} {U^{(n)}_{xi}}^\dag M^{(n)}_{ijyy'} {U^{(n)}_{jx'}}.  
\label{ren_T}
\end{align}
Note that  $x,y$ are truncated indices which run from $1$ to $\chi$ 
although $i,j$ run from $1$ to $\chi^2$.  
Fig.~\ref{fig:M_of_HOTRG} (b) shows the renormalization of the 2d HOTRG.

Thus $Z$ is again expressed as Eq.~(\ref{tensor_network}) with Eq.~(\ref{ren_T}). 
After finishing the renormalization along the $y$ direction, we move on to the renormalization along the $x$ direction.
This can be carried out by simply repeating Eqs.~(\ref{M_HOTRG})--(\ref{ren_T}) 
because the ordering of indices in  the lhs of Eq.~(\ref{ren_T})
have already changed
to be able to do that.
 The HOTRG ends in $N$ steps for a finite volume lattice of $V=2^N (=2^{\frac{N}{2}} \times 2^\frac{N}{2})$ 
 where $N$ is an even integer. 
 We finally obtain the value of $Z$ by evaluating $Z \approx \sum_{x,y} T^{(N)}_{xxyy}$ 
 where $T^{(0)}$ is an initial tensor. 

Since the isometry $U$ is determined from the local diagonalization for $K$ in Eq.~(\ref{left_unitary}),
the HOTRG also ignores the influence of environment tensors as in case of TRG. 
In Ref. \cite{HOTRG}, the second renormalization group for the HOTRG
is proposed as the HOSRG. 
In the HOSRG,
the HOTRG is used only for obtaining 
the initial set of $T^{(n)}$ and $U^{(n)}$ for $n=0,1,\ldots,N-1$. Then 
$T^{(n)}$ and $U^{(n)}$ are updated by using the recurrence relation for the environment tensors
and a diagonalization of the bond density matrix.
In this paper, we employ  
$X^{(n)}$ as described in Ref. \cite{MERAupdate}
 and defined below to update  $U^{(n)}$ instead of using the bond density matrix 
because the computational cost is reduced. 

\begin{figure}[H]
\centering
\includegraphics[width=50mm]{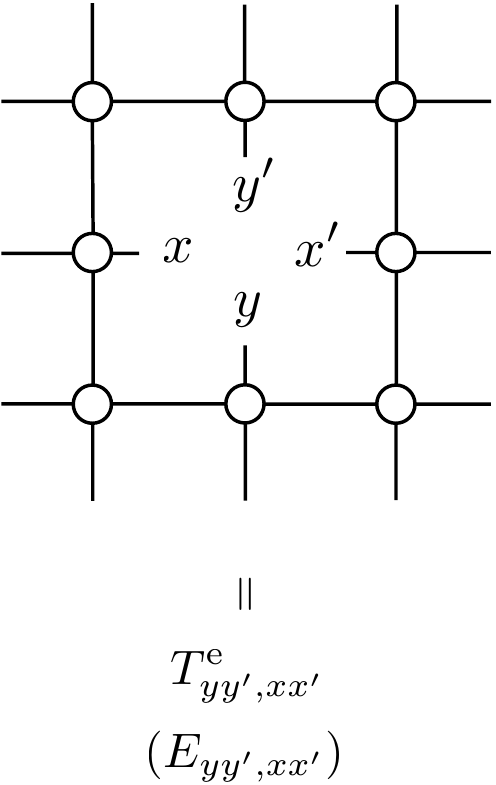}
\caption{
Environment tensor of the HOTRG. 
}
 \label{fig:env_HOTRG}
\end{figure}

\begin{figure}[H]
\centering
\includegraphics[width=160mm]{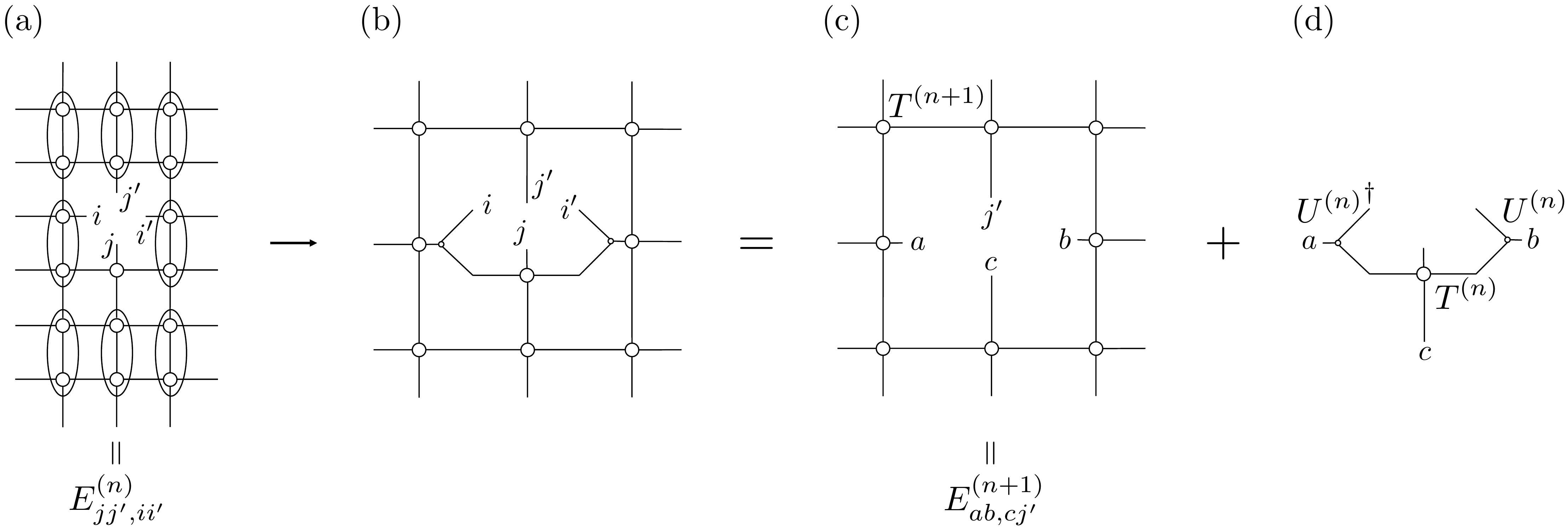}
\caption{
Renormalization of environments in the HOSRG. 
}
 \label{fig:renormalization_environments_HOSRG}
\end{figure}

\begin{figure}[H]
\centering
\includegraphics[width=70mm]{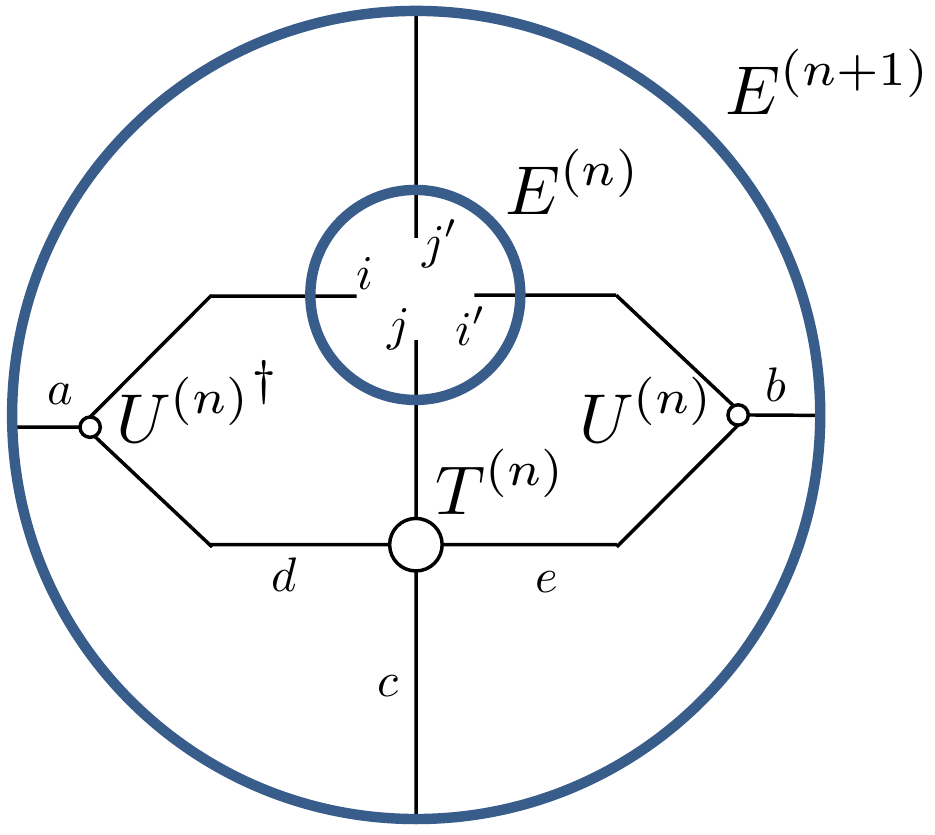}
\caption{
Schematic representation of the recurrence relation Eq.~(\ref{recurrence_relation_hosrg}). 
}
 \label{fig:reurrence_relation_hosrg}
\end{figure}

Let us express $Z$ as 
\begin{align}
Z = {\rm Tr}(T T^{\rm e}) \equiv \sum_{x,x',y,y'} T_{xx',yy'} T^{\rm e}_{yy',xx'}
\end{align}
where  $T^{\rm e}$ is the environments of $T$,  which is defined by removing $T$ from $Z$
as shown in Fig.~\ref{fig:env_HOTRG}. 
As in case of SRG, we evaluate  $T^{\rm e}$ iteratively using the HOTRG method as shown in  
Fig.~\ref{fig:renormalization_environments_HOSRG}.
For $E^{(0)}=T^{{\rm e}}$ which is the environment for $T^{(0)}$, 
the $n$-scale environment $E^{(n)}$ contains
$2^{N} - 2^n$
tensors of the scale $n$.
The environment tensors approximately satisfy a recurrence relation,
\begin{align}
  E^{(n)}_{jj',ii'} = \sum_{a,b,c,d,e} E^{(n+1)}_{ab,cj'} {U^{(n)}_{a,id}}^\dag T^{(n)}_{decj} U^{(n)}_{i'e,b},  
  \label{recurrence_relation_hosrg}
\end{align}
and we define the $i$-scale $E^{(i)}$ with the recurrence relation by solving it from 
$E^{(N)}_{ab,cd}=\delta_{ab}\delta_{cd}$. 
Fig.~\ref{fig:reurrence_relation_hosrg} shows a schematic representation of the recurrence 
relation Eq.~(\ref{recurrence_relation_hosrg}). 

\begin{figure}[H]
\centering
\includegraphics[width=85mm]{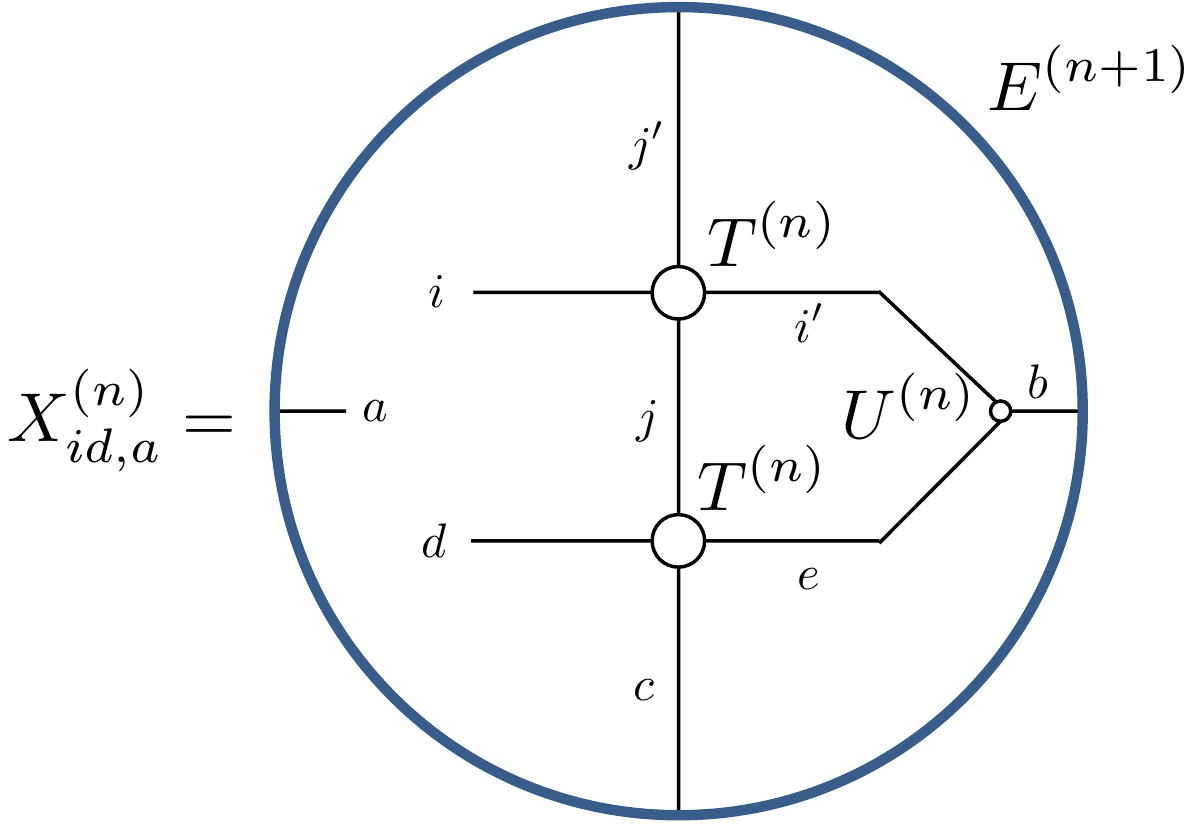}
\caption{
The graphical representation of $X^{(n)}$ of the HOSRG. 
}
 \label{fig:X_HOSRG}
\end{figure}

Once the environment
 is obtained, the isometry $U^{(n)}$ is updated from $X^{(n)}$ defined by
\begin{align}
X^{(n)}_{id,a} \equiv \sum_{i',j,j',b,c,e} E^{(n+1)}_{ab,cj'}  T^{(n)}_{ii'jj'}  T^{(n)}_{decj} U^{(n)}_{i'e,b}. 
\label{X_hosrg}
\end{align}
Fig.~\ref{fig:X_HOSRG} shows $X^{(n)}$.
We decompose $X^{(n)}$ by the SVD as 
\begin{align}
X^{(n)}_{id,a} = \sum_m u_{id,m} \sigma_m v^\dag_{ma}.
\label{x_svd}
\end{align}
Then $U^{(n)}$ is updated by
\begin{align}
U^{(n)}_{id,a} = \sum_m u_{id,m} v^\dag_{ma}. 
\label{u_update}
\end{align}
We should note here that $Z={\rm Tr }(U^{(n)\dag} X^{(n)})={U^{(n)}}^\dag_{a,id} X^{(n)}_{id,a} \approx \sum_m \sigma_m$ 
and $U^{(n)}$ is updated in a sense that $Z$ is well approximated.  

The HOSRG is given in 2d square lattice of $V=2^N$ (with the periodic boundary condition) as follows:
we compute 
all of the $T^{(n)}$ and $U^{(n)}$ ($n=0,1,\ldots,N-1$) by the HOTRG. 
Then, for these initial sets of  $T^{(n)}$ and $U^{(n)}$, 
we employ a forward-backward algorithm that repeats (i) and (ii) 
$m$ times (sweeps) where $m$ is chosen so that converged or better results are obtained
\footnote{
$m=5\sim 10$ is large enough in the actual computations of 2d Ising model on $V=2^{50}$
presented in section \ref{sec:numerical_results}. 
}:
\begin{itemize}
\item[(i)] 
$E^{(n)}$ ($n=1,2,\ldots,N$) are computed by solving the recurrence relation Eq.~(\ref{recurrence_relation_hosrg}).  
\item[(ii)]
Two procedures are repeated for $n=0,1,\ldots, N-1$: 
$U^{(n)}$ is updated by 
Eqs.~(\ref{X_hosrg})--(\ref{u_update}) from $E^{(n+1)}, T^{(n)},U^{(n)}$. Then $T^{(n+1)}$ is updated by
Eq.~(\ref{ren_T}) from $T^{(n)}$ and the new $U^{(n)}$.  
\end{itemize}
Finally we evaluate $Z \approx \sum_{x,y} T^{(N)}_{xxyy}$ in the HOSRG as well as the HOTRG.

The computational costs of HOTRG/HOSRG scale with ${\cal O}(N \chi^7)$ and 
${\cal O}(mN \chi^7)$, respectively,  
since the contraction in Eq.~(\ref{ren_T}) takes a cost of $N \chi^7$
and the recurrence relation and $X^{(n)}$ in Eq.~(\ref{X_hosrg}) are evaluated in a cost of  $m N \chi^7$.
Fig.~4 of  Ref. \cite{HOTRG}, Fig.~10 of Ref. \cite{HOTRG_pbc} and Figs.~\ref{fig:T_error_24}--\ref{fig:T_error_48} presented in section \ref{sec:numerical_results} show
that the HOSRG significantly improves the accuracy of results.  
%

%%%%%%%%%%%%%%%%%%%%%%%%%%%%%%%%%%%%%%%%%%%%%%%%%%%%%%%%%
%
%
%         3.   The triad SRG
%  
%
%%%%%%%%%%%%%%%%%%%%%%%%%%%%%%%%%%%%%%%%%%%%%%%%%%%%%%%%%

\section{Triad second renormalization group\label{sec:Triad_SRG}}
\subsection{The triad TRG in two dimensions\label{sec:Triad_TRG}}
We consider a lattice model with local interactions on $d$ dimensional hyper cubic lattice. 
The partition function of a translational invariant theory is given by a homogeneous tensor network of a rank-$2d$ tensor $T$. 
Then $T$ is naturally obtained as a polyadic decomposition: 
\begin{align}
  T_{i_1 i_2\cdots i_{2d}} = \sum_{r}
  W^{(1)}_{r i_1}W^{(2)}_{r i_2} \cdots W^{(2d)}_{r i_{2d}}, 
  \label{pd}
\end{align}
since $2d$ hopping terms stemmed from a site provide $2d$ $W_{r j}$ factors.

A triad representation of $T$ is hidden in Eq.~(\ref{pd}).
For instance, in 3 dimensions, 
we have
\begin{align}
  T_{ijklmn} = \sum_{a,b,c} A_{ija} B_{akb} C_{blc} D_{cmn}, 
  \label{3d_triad}
\end{align}
where 
\begin{align}
& A_{ija} = W^{(1)}_{ai} W^{(2)}_{aj}, \\
& B_{akb} = \delta_{ab} W^{(3)}_{ak}, \\
& C_{blc} = \delta_{bc} W^{(4)}_{bl}, \\
& D_{cmn} = W^{(5)}_{cm} W^{(6)}_{cn}.
\end{align}
Note that this kind of representation is not unique because 
we can interchange $W^{(i)} \leftrightarrow W^{(j)}$ for $A,B,C,D$. 
In Ref. \cite{Kadoh:2019kqk}, the computational cost is reduced by 
applying the HOTRG-like renormalization to the triad tensor network because all of the tensors 
are made of rank-$3$ tensors in the triad tensor network.

\begin{figure}[H]
\centering
\includegraphics[width=160mm]{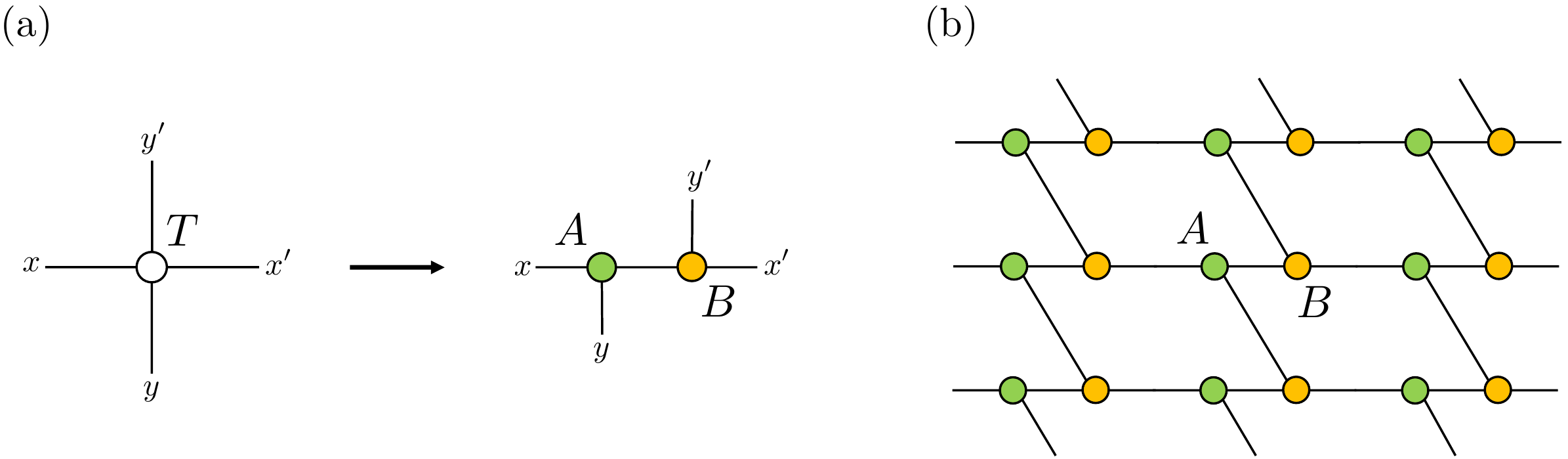}
\caption{
(a) Triad representation of a tensor and (b) 
the triad network given by Eq.~(\ref{triad}). 
}
 \label{fig:triad_network}
\end{figure}

Let us consider two dimensional case of the triad representation like Eq.~(\ref{3d_triad}) 
as 
\begin{align}
  T_{xx'yy'} = \sum_{a} A_{xya}B_{ay'x'},
  \label{triad}
\end{align}
where
\begin{align}
  A_{xya} &= W^{(1)}_{a x} W^{(2)}_{a y}, \quad 
  B_{ayx} = W^{(3)}_{a y} W^{(4)}_{a x}.
  \label{AB_initial_triad}
\end{align}
The tensor network of Eq.~(\ref{triad}) is introduced by setting $T$ of Eq.~(\ref{tensor_network}) to Eq.~(\ref{triad}). 
Fig.~\ref{fig:triad_network} shows the triad representation and its tensor network, respectively. 
In two dimensions, the triad network with the periodic boundary condition
can be regarded as one on a honeycomb lattice with unusual boundary condition. 
\footnote{
In Ref. \cite{Zhao2010}, the tensor network on a square lattice is embedded into 
one on a honeycomb lattice with an enlarged intermediate index that runs from $1$ to $\chi^2$.  
}

The triad TRG in two dimensions is simply introduced by setting the rank-$4$ tensor $T$ 
to Eq.~(\ref{triad}) 
in the HOTRG algorithm. 
The renormalized tensor given by Eq.~(\ref{ren_T}) can be decomposed into 
renormalized triads by the SVD as
\begin{align}
T^{(n+1)}_{yy'xx'} \approx \sum_{a=1}^\chi A^{(n+1)}_{yxa} B^{(n+1)}_{ax'y'}.  
\label{ren_triads_D6}
\end{align}
Thus the tensor renormalization is realized on the triad network.
The cost of contracting indices in Eq.~(\ref{ren_T}) is reduced to ${\cal O}(\chi^6)$ using  Eq.~(\ref{triad})  for $M^{(n)}$.
Note that other parts such as creating the isometry $U^{(n)}$ shown in Eq.~(\ref{left_unitary}) and the SVD of Eq.~(\ref{ren_triads_D6}) scales 
with $\chi^6$ without Eq.~(\ref{triad}). 
With the help of the randomized SVD (RSVD), 
the cost is further reduced to $\chi^5$.
Thus the cost of triad TRG method in two dimensions scales with ${\cal O} (N \chi^6)$ 
(or ${\cal O} (N \chi^5)$ with the RSVD).

\subsection{A preliminary algorithm of the triad SRG}

We introduce a preliminary SRG algorithm on a triad network 
before the triad SRG is defined in the next section.
This algorithm is simply given 
by setting $T$ in the HOSRG to Eq.~(\ref{triad}).
The cost of the HOSRG is then reduced to  ${\cal O}(\chi^6)$ as seen below. 

The leading cost of the HOSRG, which is $\chi^7$, comes from 
three parts of its algorithm: the HOTRG in the first round, the evaluation of the environment tensors 
solving the recurrence relation in Eq.~(\ref{recurrence_relation_hosrg}) and update of isometry from  
$X^{(n)}$ in Eq.~(\ref{X_hosrg}). 
Instead of the HOTRG, we use the triad TRG 
to give the initial set of tensors $A^{(n)},B^{(n)}$ and isometries $U^{(n)}$, 
and the cost of this part is reduced to $\chi^6$.   

\begin{figure}[H]
\centering
\includegraphics[width=110mm]{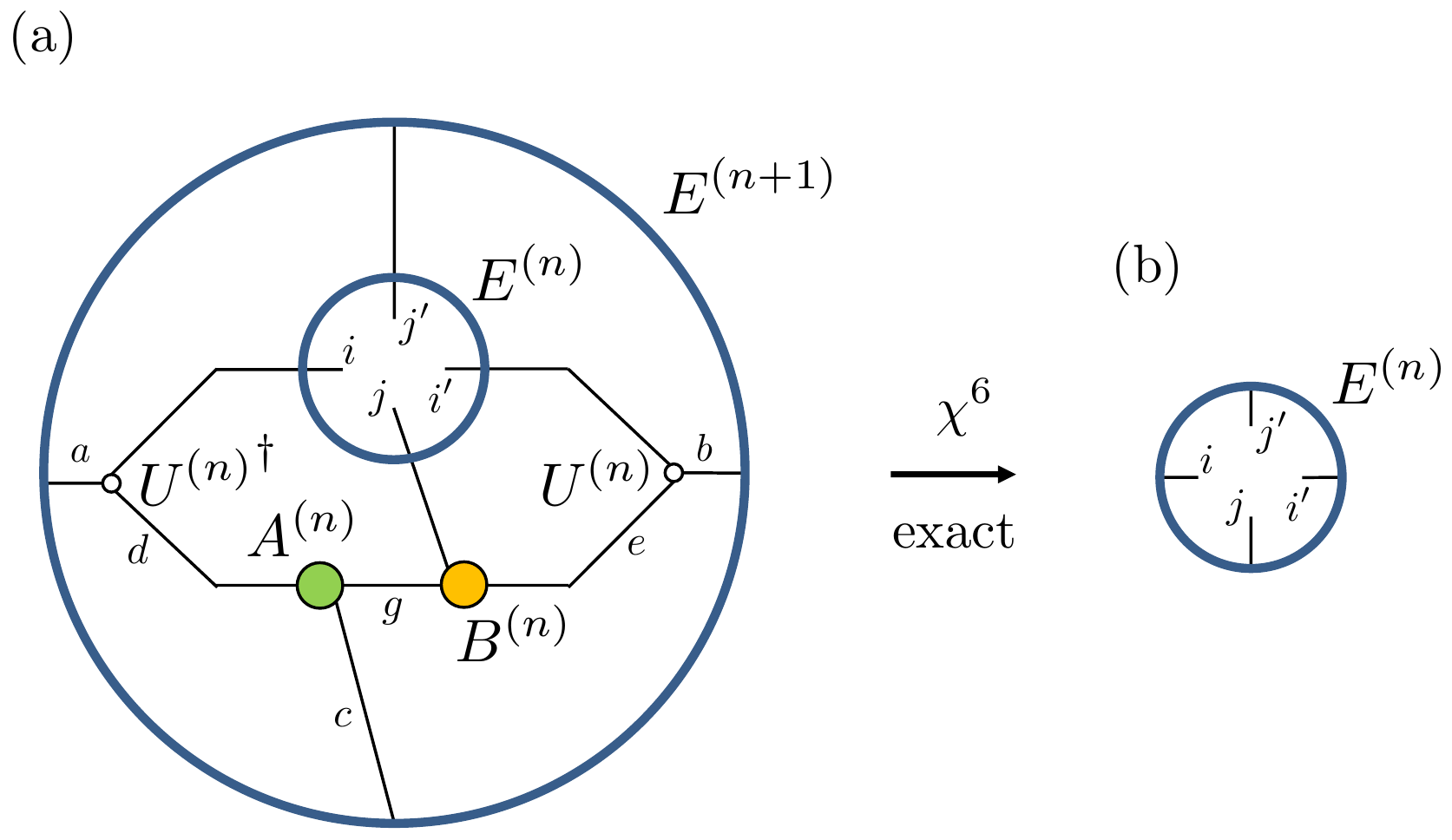}
\caption{
Recurrence relation in triad SRG ($\chi^6$) given in Eq.~(\ref{recurrence_relation_triadsrg_D6}). 
}
 \label{fig:reurrence_relation_triadsrg_D6}
\end{figure}

The recurrence relation with Eq.~(\ref{triad})
is given by 
\begin{align}
  E^{(n)}_{jj',ii'} = \sum_{a,b,c,d,e,g} E^{(n+1)}_{ab,cj'} {U^{(n)}_{a,id}}^\dag 
  A^{(n)}_{dcg}B^{(n)}_{gje}
  U^{(n)}_{i'e,b},  
  \label{recurrence_relation_triadsrg_D6}
\end{align}
where $E^{(N)}_{ab,cd}=\delta_{ab}\delta_{cd}$. 
Fig.~\ref{fig:reurrence_relation_triadsrg_D6} (a) shows the recurrence relation
of Eq.~(\ref{recurrence_relation_triadsrg_D6}). 
Contracting links in the order of 
$(d,e) \rightarrow (a,c) \rightarrow (b,g)$
leads to Fig.~\ref{fig:reurrence_relation_triadsrg_D6} (b), and these contractions take at most a cost of $\chi^6$. 
 Thus we find that the cost of solving the recurrence relation scales with  $\chi^6$.    

\begin{figure}[H]
\centering
\includegraphics[width=100mm]{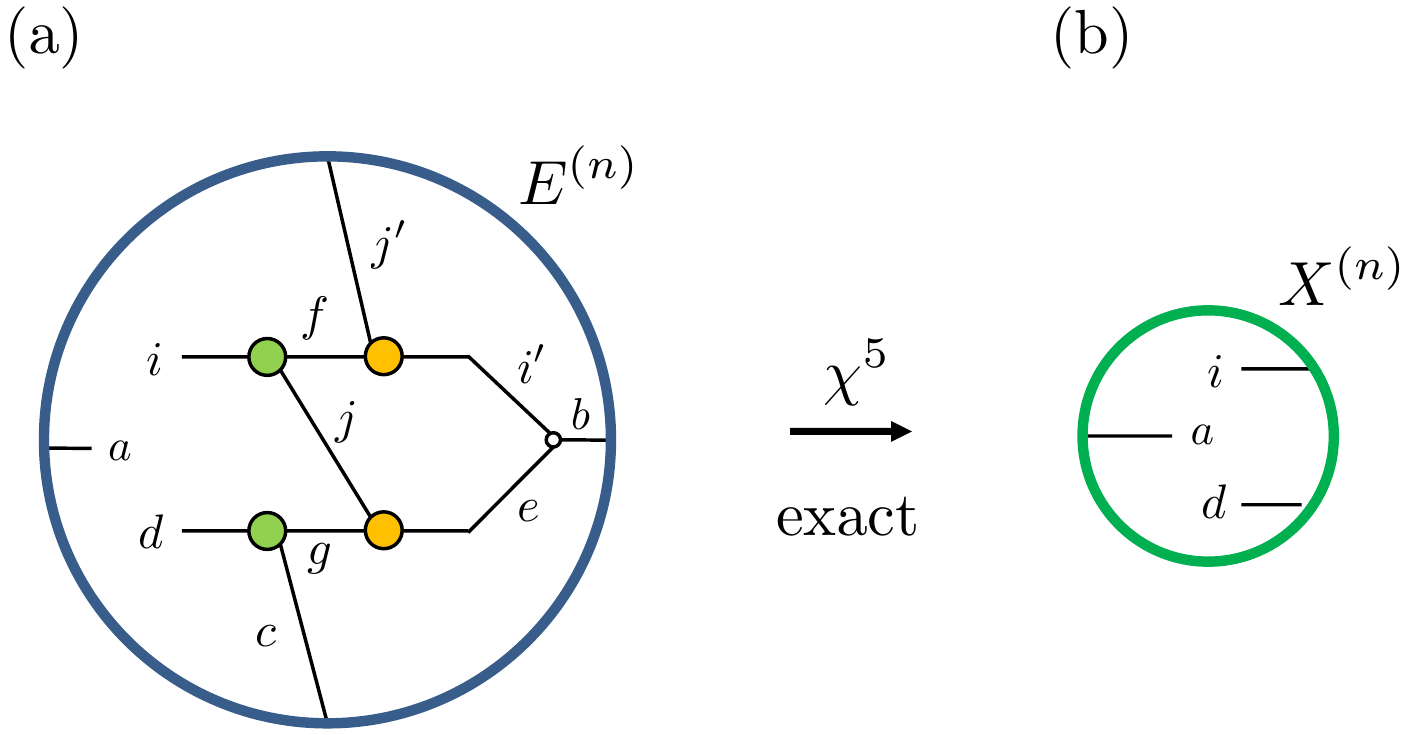}
\caption{
$X^{(n)}$ of the triad SRG ($\chi^6$). 
}
 \label{fig:X_triadSRG_D6}
\end{figure}

Fig.~\ref{fig:X_triadSRG_D6} shows $X^{(n)}$ in which $T$
is replaced by Eq.~(\ref{triad}) as
\begin{align}
X^{(n)}_{id,a} \equiv \sum_{i',j,j',b,c,e,f,g} E^{(n+1)}_{ab,cj'}  
A^{(n)}_{ijf} B^{(n)}_{fj'i'} A^{(n)}_{dcg} B^{(n)}_{gje}
U^{(n)}_{i'e,b}. 
\label{X_triadsrg_D6}
\end{align}  
Fig.~\ref{fig:X_triadSRG_D6} (b) is obtained from Fig.~\ref{fig:X_triadSRG_D6} (a)  
by contracting links in the order of $(g,i') \rightarrow (j,j',b) \rightarrow (c,e,f)$.
Thus we can evaluate $X^{(n)}$ within the cost of $\chi^6$. 
Once $X^{(n)}$ is obtained, $U^{(n)}$ is updated by Eqs.~(\ref{x_svd}) and (\ref{u_update}) 
in the cost of $\chi^4$. 

The renormalized rank-$4$ tensor $T^{(n+1)}$ is given by the updated $U^{(n)}$ 
according to Eq.~(\ref{ren_T}).  
Once $T^{(n+1)}$ and its environment $E^{(n+1)}$ 
are given, renormalized triads $A^{(n+1)}$ and $B^{(n+1)}$ are obtained from 
the fundamental procedure of SRG presented in 
Eqs.~(\ref{def:tildeM})--(\ref{good_svd_M}). Setting  $T^{(n+1)},E^{(n+1)}$ to $M,M^{\rm e}$, respectively,
we find a triad representation of the next step: 
\begin{align}
T^{(n+1)}_{ijkl} \approx \sum_{a=1}^\chi A^{(n+1)}_{ika} B^{(n+1)}_{alj}
\label{new_triad_D6}
\end{align}
where the definitions of $A$ and $B$ can be read from $C$ and $D$ of Eq.~(\ref{good_svd_M}). 
This procedure scales with $\chi^6$ as it consists of the matrix products 
and the SVDs for $\chi^2 \times \chi^2$ matrices.

The preliminary algorithm  in 2d square lattice of $V=2^N$ with PBC 
is summarized below. 
The initial set of $A^{(n)}, B^{(n)}$ and $U^{(n)}$ ($n=0,1,\ldots,N-1$) are generated by the triad TRG. 
We then repeat (i) and (ii) $m$ times (sweeps):
\begin{itemize}
\item[(i)] 
$E^{(n)}$ ($n=1,2,\ldots,N$) are computed by solving the recurrence relation Eq.~(\ref{recurrence_relation_triadsrg_D6}). 
\item[(ii)] Two procedures are repeated for $n=0,1,\ldots, N-1$: 
$U^{(n)}$ is updated by using $X^{(n)}$ in 
Eq.~(\ref{X_triadsrg_D6}) and its SVD shown as Eqs.~(\ref{x_svd}) and (\ref{u_update}) 
from $E^{(n+1)}, A^{(n)}, B^{(n)},U^{(n)}$. Then 
 $A^{(n+1)}, B^{(n+1)}$ are updated by
Eqs.~(\ref{ren_T}) and (\ref{new_triad_D6})  from  $E^{(n+1)}, A^{(n)}, B^{(n)}$  and the new $U^{(n)}$.  
\end{itemize}
where $m$ is chosen so that converged or better results are obtained. 
We finally evaluate  
$Z \approx \sum_{i,j,a} A^{(N)}_{ija} B^{(N)}_{aji}$, or 
$Z \approx \sum_{i,j} T^{(N)}_{iijj}$
with $T^{(N)}$ in the last step (ii).  
The cost of this algorithm scales with ${\cal O} (m N \chi^6)$.

\subsection{The triad SRG}

The preliminary algorithm shown in the previous section 
was obtained by setting $T$ to the triad representation Eq.~(\ref{triad}) in the HOSRG.
We define the triad SRG further decomposing environment and intermediate 
tensors with the randomized SVD(RSVD) (See, for example, Refs.\cite{RSVD,RSVDTRG} for the details of RSVD). 
The algorithm proposed here scales with $\chi^5$.

The $\chi^6$ dependence on the cost of the preliminary algorithm is originated from four parts: 
(i) the triad TRG in the first round,
(ii) the creation of the environment tensors with  the recurrence relation Eq.~(\ref{recurrence_relation_triadsrg_D6}),
(iii) the update of isometry with $X^{(n)}$ given by Eq.~(\ref{X_triadsrg_D6}),
(iv) the reconstruction of the triads shown in Eq.~(\ref{new_triad_D6}).
With the RSVD, the cost of  (i) is reduced to $\chi^5$ as described in Ref. \cite{Kadoh:2019kqk}. 
Decomposing the environment tensor reduces the cost of (ii) to $\chi^5$.
The cost of (iii) is also reduced to $\chi^5$ by decomposing an intermediate 
tensor ${\cal M}$ defined later. 
For (iv),  with the decomposed ${\cal M}$, 
the renormalized triads are obtained at a cost of  $\chi^5$.
We see these points in detail below.

\begin{figure}[H]
\centering
\includegraphics[width=150mm]{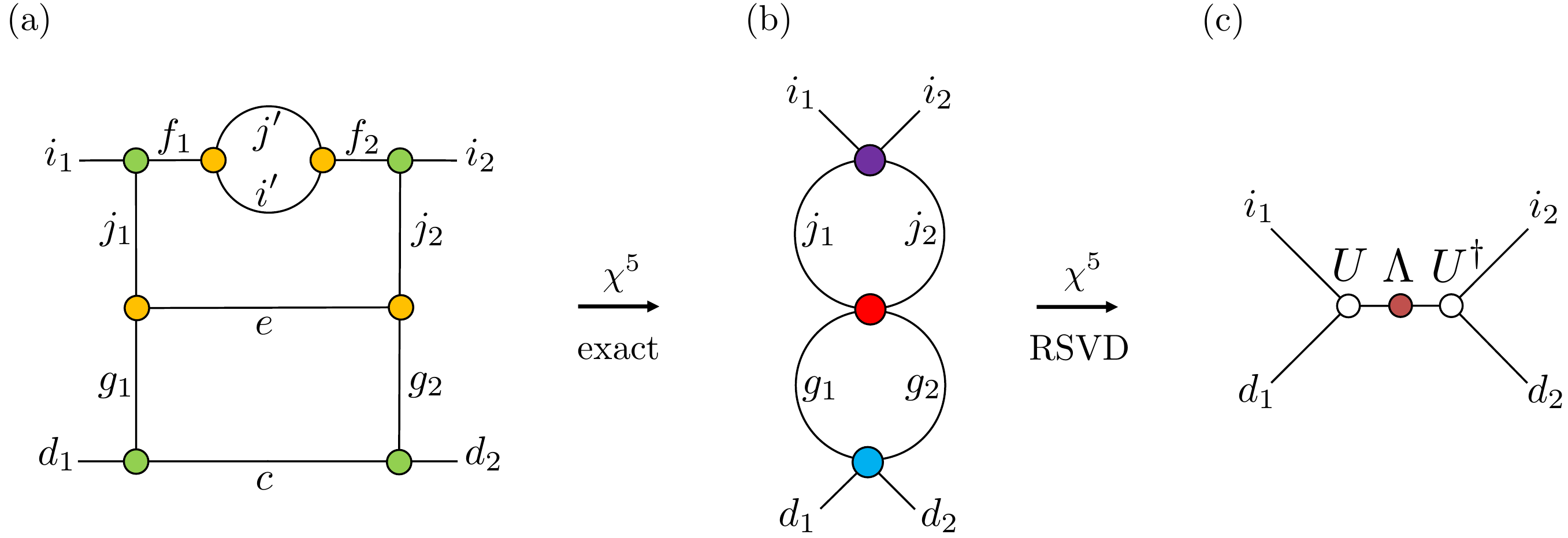}
\caption{
Decomposition of $K$: 
(a) $K$ with the triads. (b) is obtained by contracting six indices. 
(c) is the diagonalization of $K$ with the RSVD.   
}
 \label{fig:K_triadsrg_D5}
\end{figure}

\begin{figure}[H]
\centering
\includegraphics[width=150mm]{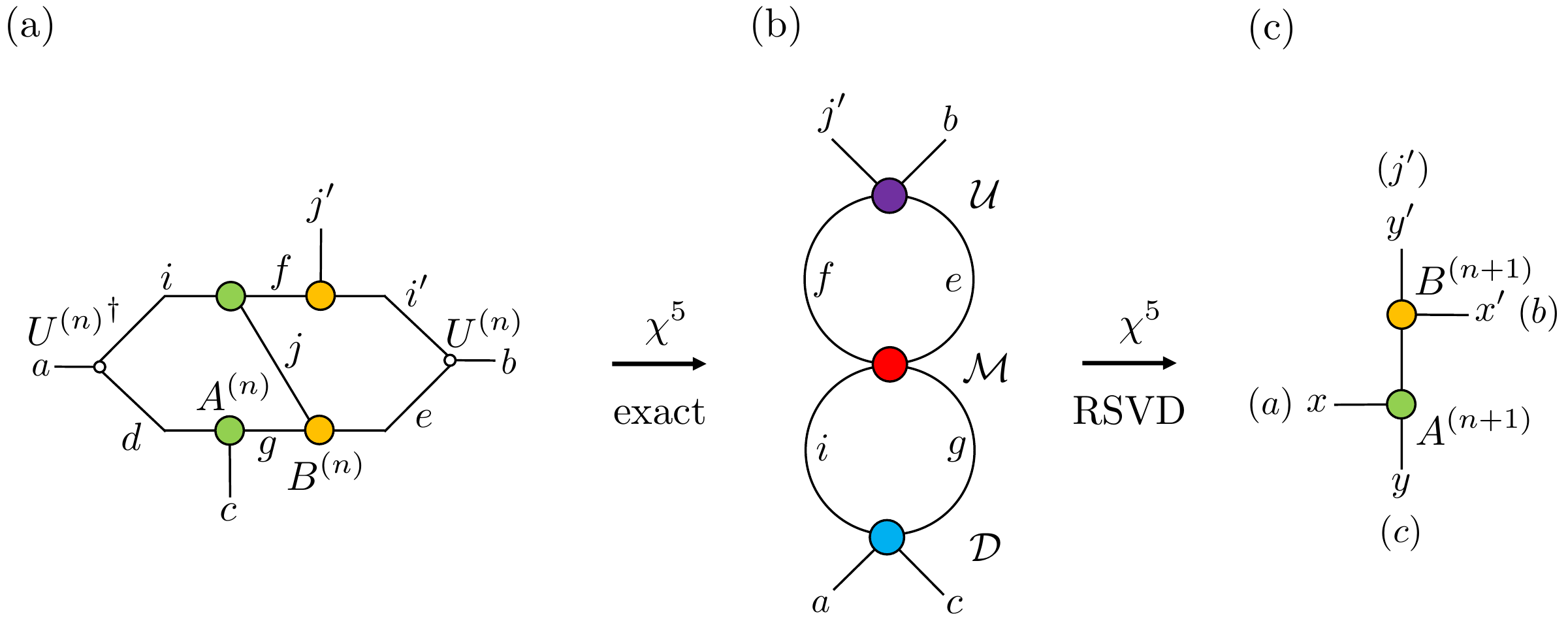}
\caption{
Renormalization of triads: 
(a) four triads with two isometries.
(b) is obtained by contracting the indices $d,j,i'$. 
(c) is the RSVD of (b).   
}
 \label{fig:ren_triads_triadsrg_D5}
\end{figure}

In the triad TRG, the leading cost comes
from creating the isometry $U^{(n)}$ and the renormalized triads $A^{(n+1)}$ and $B^{(n+1)}$
(See Eqs.~(\ref{M_HOTRG}), (\ref{left_unitary}), (\ref{ren_T}) with Eqs.~(\ref{triad}) and (\ref{new_triad_D6})).
Fig.~\ref{fig:K_triadsrg_D5} (a) shows $K=M^\dag M$ with the triads.
A two-loop graph with external four legs shown in 
Fig.~\ref{fig:K_triadsrg_D5} (b) is obtained in a cost of $\chi^5$ by contracting indices $a,a',x_1',x_2',y,y'$ from (a).
We can diagonalize the two-loop graph to create the isometry $U^{(n)}$ 
within $\chi^5$ by using the randomized tricks  such as the RSVD  twice. 
See the appendix Ref. \cite{Kadoh:2019kqk} for more details.   
Using $U^{(n)}$, the renormalized triads 
are obtained
as in Fig.~\ref{fig:ren_triads_triadsrg_D5}.  We again encounter a two loop graph shown in Fig.~\ref{fig:ren_triads_triadsrg_D5} (b)    
by contracting indices $i',j,d$
where 
\begin{align}
 & {\cal U}^{(n)}_{fe, j'b}= \sum_{i'} B^{(n)}_{fj'i'} U^{(n)}_{i'e,b},
  \label{calU} \\
 & {\cal M}^{(n)}_{ig,fe}= \sum_j A^{(n)}_{ijf} B^{(n)}_{gje} ,
 \label{calM}\\
  & {\cal D}^{(n)}_{ac,ig}= \sum_{d} U^{(n)\dag}_{a,id} A^{(n)}_{dcg}.  \label{calD}
\end{align}
The single use of RSVD again reduces the cost of decomposing  Fig.~\ref{fig:ren_triads_triadsrg_D5} (b) 
into Fig.~\ref{fig:ren_triads_triadsrg_D5} (c).
Thus the cost of triad TRG 
becomes $\chi^5$. 

\begin{figure}[H]
\centering
\includegraphics[width=150mm]{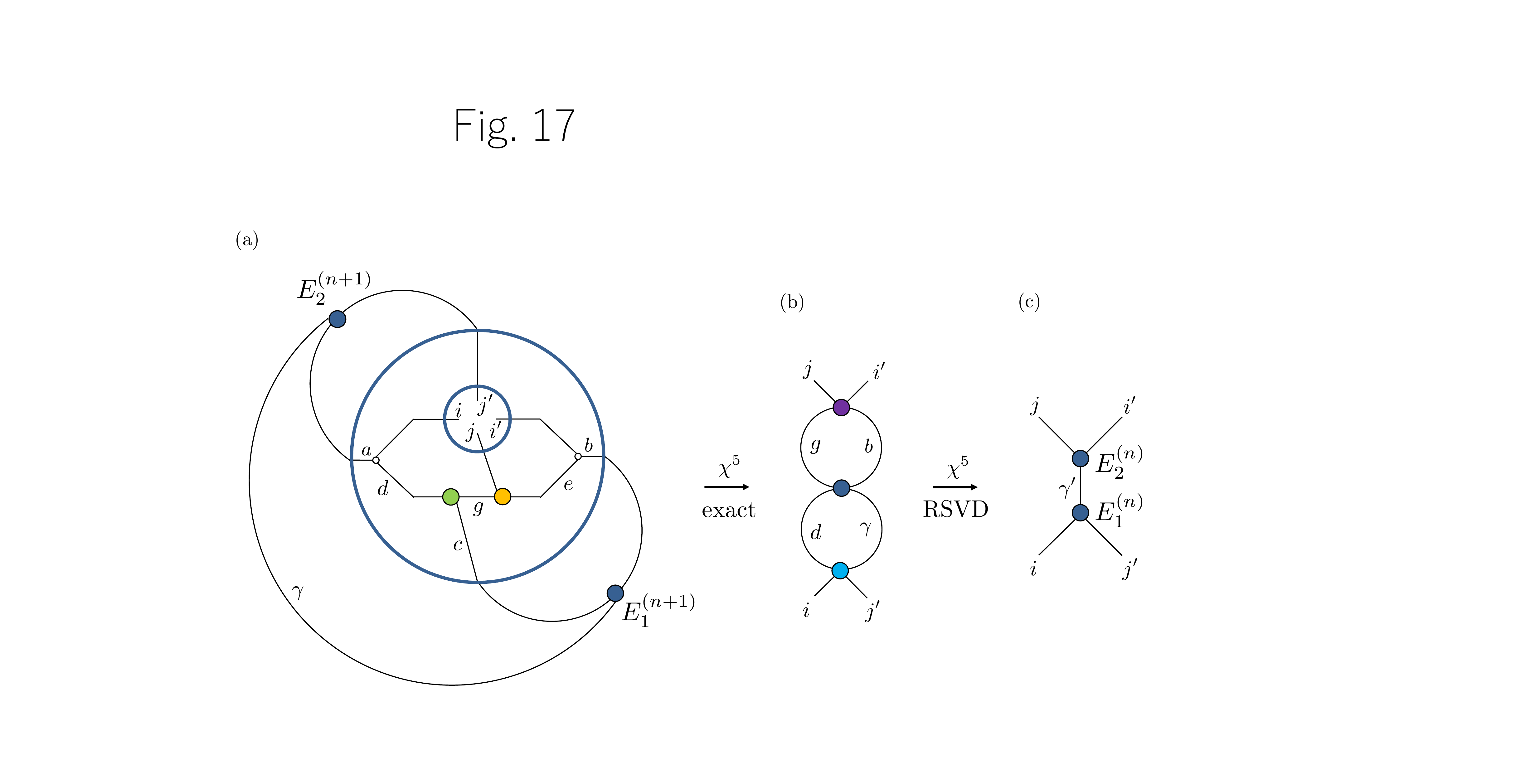}
\caption{
Recurrence relation in the triad SRG ($\chi^5$) given in Eq.~(\ref{recurrence_relation_triadsrg_D6}) with Eq.~(\ref{triad_env}). 
}
 \label{fig:reurrence_relation_triadsrg_D5}
\end{figure}

The rank-$4$ environment tensor $E^{(n)}$ is solved iteratively by
using the recurrence relation Eq.~(\ref{recurrence_relation_triadsrg_D6}).
With the RSVD, $E^{(n)}$ is obtained as a product 
of two rank-$3$ tensors 
$E_1^{(n)}$ and $E_2^{(n)}$: 
\begin{align}
E^{(n)}_{jj',ii'} \approx \sum_{\delta^\prime=1}^\chi E^{(n)}_{2,ji'\delta^\prime}  E^{(n)}_{1,j'i\delta^\prime}  \qquad {\rm for} \ \ 0 \le n< N.
\label{triad_env}
\end{align}
The cost of
computing the environment tensors
is then reduced to $\chi^5$ by 
employing Eq.~(\ref{triad_env}) in its rhs.  
Fig.~\ref{fig:reurrence_relation_triadsrg_D5} shows how to obtain a triad form of the environment in  Eq.~(\ref{triad_env}) using Eq.~(\ref{recurrence_relation_triadsrg_D6}).
Thank to Eq.~(\ref{triad_env}), a two loop graph 
shown in Fig.~\ref{fig:reurrence_relation_triadsrg_D5} (b) is obtained by contracting indices 
$a,c,e$ of Fig.~\ref{fig:reurrence_relation_triadsrg_D5} (a). 
\footnote{For $n =N$, a single loop graph actually appears. 
Fig.~\ref{fig:reurrence_relation_triadsrg_D5} (c) is also obtained at the same cost in that case.  
}
For such a multi loop graph, the RSVD is employed to obtain Fig.~\ref{fig:reurrence_relation_triadsrg_D5} (c)
 in a cost of $\chi^5$. 
We can obtain  $E^{(n)}$ of rank-$4$ as 
its rank-$3$ representation Eq.~(\ref{triad_env}) within a cost of $\chi^5$. 

\begin{figure}[H]
\centering
\includegraphics[width=15mm]{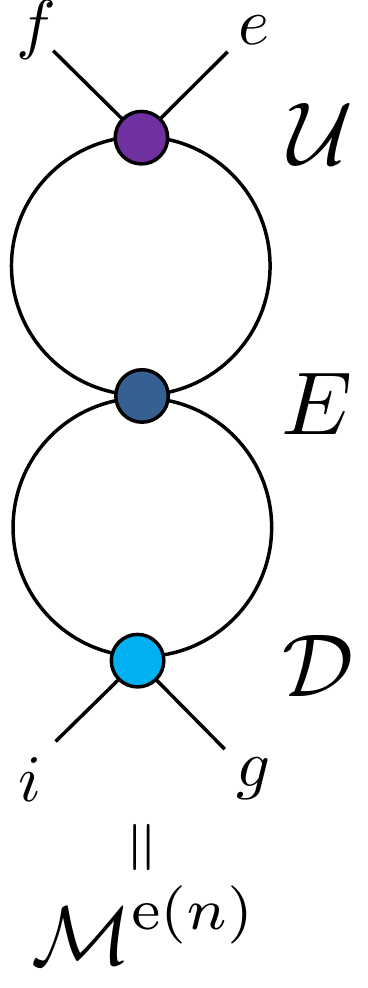}
\caption{
$ {\cal M}^{{\rm e}\,(n)}$ given by Eq.~(\ref{Me_triadSRG}). 
}
 \label{fig:Me_triadsrg_D5}
\end{figure}

\begin{figure}[H]
\centering
\includegraphics[width=160mm]{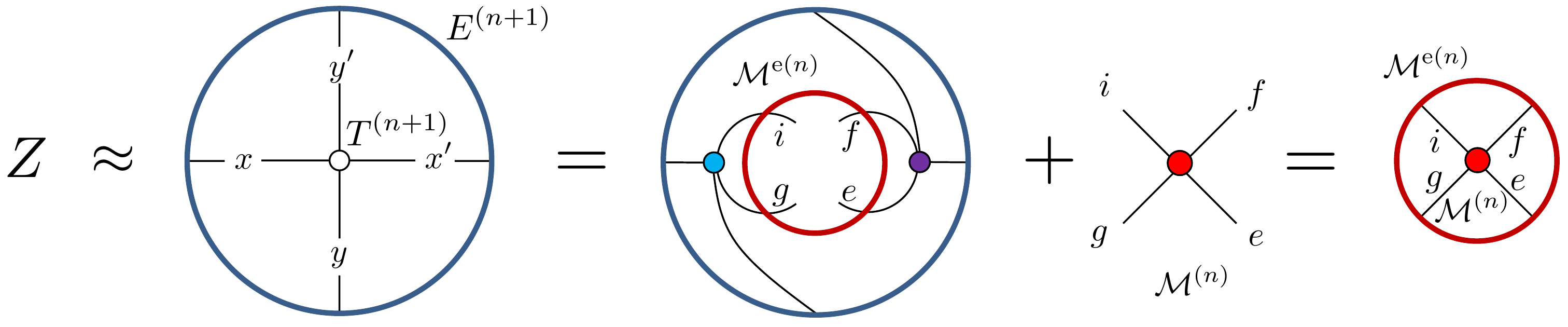}
\caption{
Partition function in terms of ${\cal M}$. 
}
 \label{fig:Z_in_M}
\end{figure}

To create the isometry in lower cost, we decompose ${\cal M}^{(n)}$ given by Eq.~(\ref{calM})
using the SRG procedure.
\footnote{
One may locally decompose ${\cal M}$, but in that case we found that accuracy of the free energy becomes worse.
}
 The partition function may be expressed as 
\begin{align}
Z \approx {\rm Tr} ({\cal M}^{(n)} { \cal M}^{{\rm e}\, (n)}) \equiv  \sum_{i,g,f,e} {\cal M}^{(n)}_{ig,fe}{ \cal M}^{{\rm e}\, (n)}_{fe,ig}.
\end{align}
Since  $Z \approx {\rm Tr} (T^{(n)} T^{{\rm e}(n)})$ at $n$th renormalization step with $E^{(n)}=T^{{\rm e}(n)}$,  we may define 
${ \cal M}^{{\rm e}\,(n)}$ as
\begin{align}
{ \cal M}^{{\rm e}\,(n)}_{fe,ig} =  {\cal U}^{(n)}_{fe, y'x'} E^{(n+1)}_{xx',yy'}   {\cal D}^{(n)}_{xy,ig}. 
\label{Me_triadSRG}
\end{align}
Figs.~\ref{fig:Me_triadsrg_D5} and \ref{fig:Z_in_M} show ${ \cal M}^{{\rm e}\,(n)}$
and 
 a clear relationship between $(T^{(n+1)},T^{{\rm e} (n+1)})$ and 
$({\cal M}^{(n)}, { \cal M}^{{\rm e}\,(n)})$.

We employ the SRG procedure given in Eqs. (\ref{def:tildeM})--(\ref{good_svd_M}) with the RSVD to decompose ${\cal M}$. 
${\cal M}^{(n)}$ and $ {\cal M}^{{\rm e}\,(n)}$ are set to 
$M$ and $M^{\rm e}$, respectively. We decompose $M^{\rm e}$ as 
$M^{\rm e} \approx U_{\rm e} \Lambda'_{\rm e} V_{\rm e}^\dag$
in a similar way to the RSVD but with a thick QR decomposition and a thick SVD
with a cost of $\chi^5$. 
The decomposition procedure is applied directly to the matrix product form of
${\cal M}^{{\rm e}\,(n)}$ shown in Fig.~\ref{fig:Me_triadsrg_D5}. 
Then $U_{\rm e}$ and $V_{\rm e}$ are obtained as $\chi^2 \times \chi^2$ unitary matrices, 
while $\Lambda'_{\rm e}$ is a $p\chi \times p\chi$ diagonal matrix in which singular values $\sigma_i$ are sorted in the descending order.
\footnote{
$p$ is a tunable parameter, which is taken to be O(1) so that the leading cost is kept. In our implementation, $p$ is set to be 2.
}
We enlarge $\Lambda'_{\rm e}$ to a $\chi^2 \times \chi^2$ diagonal matrix $\Lambda_{\rm e}$ by supplementing $\epsilon$ 
in its diagonal elements as \footnote{
We need the $\chi^2 \times \chi^2$ matrix $\Lambda_{\rm e}^{-1}$, not $\chi \times \chi$ one, to read $M$ from $\tilde M$. 
}
\begin{align}
\Lambda_{\rm e} = \left(
\begin{array}{cc}
\Lambda'_{\rm e} &   0 \\
     0         & \epsilon \mathbf{1} \\ 
\end{array}
\right)
\end{align}
where $\epsilon=\sigma_1 \times 10^{-15}$. Thus we have $M^{\rm e} \approx U_{\rm e} \Lambda_{\rm e} V_{\rm e}^\dag$
where $U_{\rm e},V_{\rm e},\Lambda_{\rm e}$ are $\chi^2 \times \chi^2$ matrices. 
The RSVD is again applied directly to a matrix product $\tilde M$ defined by Eq.~(\ref{def:tildeM}),
and we obtain $\tilde M \approx \tilde U \tilde \Lambda \tilde V^\dag$ in a cost of $\chi^5$. 
Thus
we have the global decomposition of $\M$
in Eq.~(\ref{good_svd_M})
as
\begin{align}
{\cal M}^{(n)}_{ig,fe} \approx \sum_{\delta=1}^\chi C^{(n)}_{ig,\delta} D^{(n)}_{fe,\delta}
\label{svd_M_triadsrg_D5}
\end{align}
where  $C^{(n)}_{ij,m}= (V_{\rm e} \sqrt{\Lambda_{\rm e}}^{-1} \tilde U \sqrt{\tilde \Lambda})_{ij,m}$
and  $D^{(n)}_{ij,m} =  ( \sqrt{\tilde \Lambda} \tilde V^\dag  \sqrt{\Lambda_{\rm e}}^{-1} U_{\rm e}^\dag )_{m,ij}$, 
which are computed within $\chi^5$.

\begin{figure}[H]
\centering
\includegraphics[width=100mm]{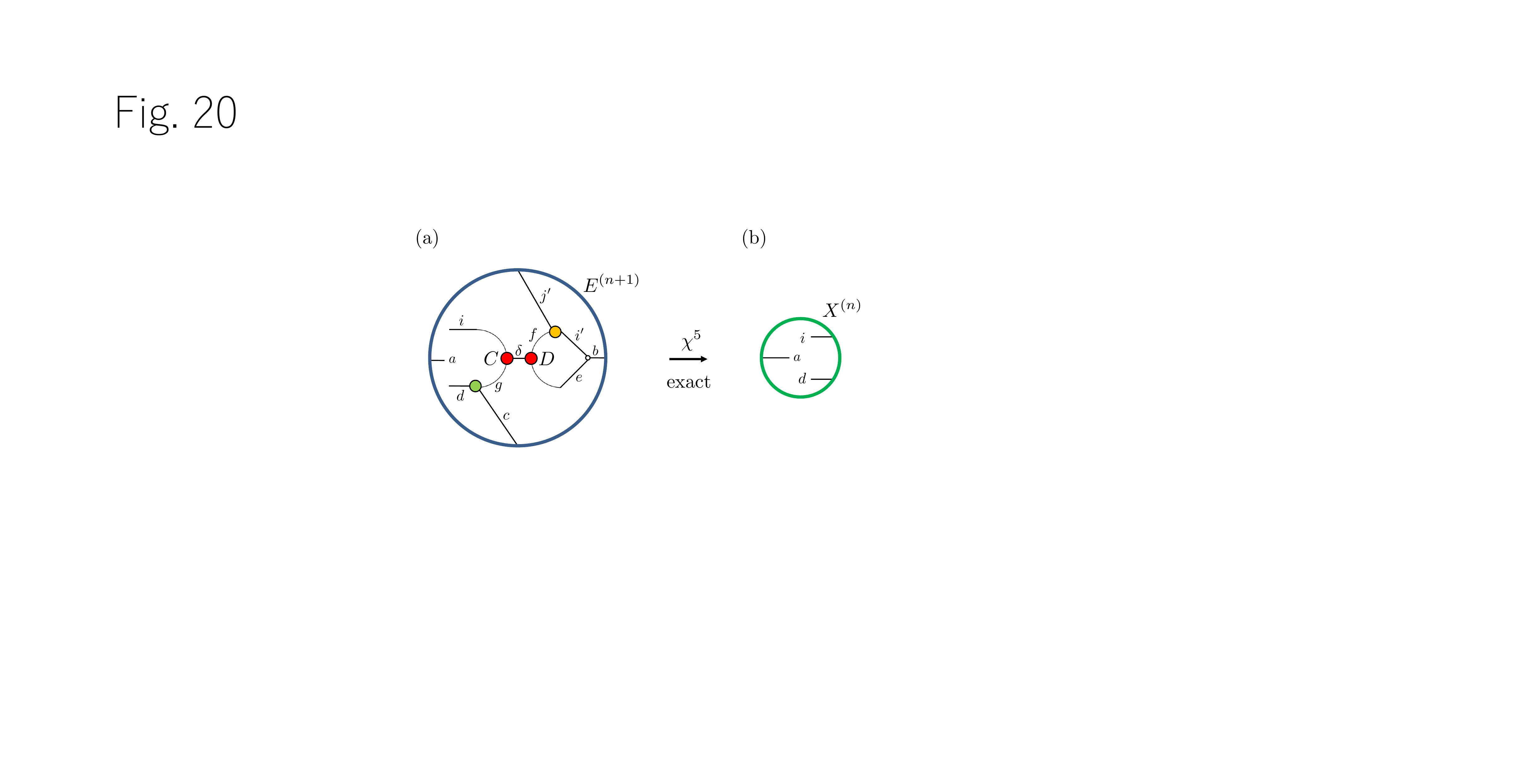}
\caption{
$X^{(n)}$ of the triad SRG ($\chi^5$). 
}
 \label{fig:X_triadSRG_D5}
\end{figure}

$X^{(n)}$ is defined with the decomposed ${\cal M}$ as
\begin{align}
X^{(n)}_{id,a} \equiv \sum_{i',j',b,c,e,f,g,\delta} E^{(n+1)}_{ab,cj'}  
 B^{(n)}_{fj'i'} A^{(n)}_{dcg} 
  C^{(n)}_{ig,\delta} D^{(n)}_{fe,\delta}
U^{(n)}_{i'e,b}. 
\label{X_triadsrg_D5}
\end{align}  
Note that Eq.~(\ref{X_triadsrg_D5})
and
Eq.~(\ref{X_triadsrg_D6})
are the same but $A$ and $B$ of Eq.~(\ref{X_triadsrg_D6}),
which correspond to $\M$, are
replaced by Eq.~(\ref{svd_M_triadsrg_D5}).
Fig.~\ref{fig:X_triadSRG_D5} shows
a graphical representation of $X^{(n)}$ given by Eq.~(\ref{X_triadsrg_D5}). 
Fig.~\ref{fig:X_triadSRG_D5} (b) is obtained in a cost of $\chi^5$ 
from Fig.~\ref{fig:X_triadSRG_D5} (a) by contracting 
indices in the order of $(f,e,i') \rightarrow (b,j',g) \rightarrow (c,\delta)$. 
Once $X^{(n)}$ is obtained, the isometry $U^{(n)}$ is updated by Eqs. (\ref{x_svd}) and (\ref{u_update}). 

With the updated isometry $U^{(n)}$ and the decomposed ${\cal M}^{(n)}$,
the renormalized triads are simply 
defined by
\begin{align}
& A^{(n+1)}_{yx\delta} =  \sum_{i,g} {\cal D}^{(n)}_{xy,ig} C^{(n)}_{ig,\delta},
\label{new_triad_A}
\\ 
& B^{(n+1)}_{\delta x'y'} =  \sum_{f,e}  D^{(n)}_{fe,\delta} {\cal U}^{(n)}_{fe,y'x'}, 
\label{new_triad_B}
\end{align}
where $C^{(n)},D^{(n)}$ are given by Eq.~(\ref{svd_M_triadsrg_D5}),  
and ${\cal U}^{(n)}, {\cal D}^{(n)}$ are given by  Eqs. (\ref{calU}) 
and  (\ref{calD}) in which $U^{(n)}$ is updated by Eq.~(\ref{x_svd}), (\ref{u_update}) and (\ref{X_triadsrg_D5}).

The triad SRG is thus given in 2d square lattice of $V=2^N$ with PBC as follows. 
We first generate the initial set of 
 $A^{(n)}, B^{(n)}$ and $U^{(n)}$ ($n=0,1,\ldots,N-1$) by the triad TRG
 with the $\chi^5$ algorithm reviewed above in this subsection. 
We then repeat (i) and (ii) 
$m$ times (sweeps): % for $i=0,1\ldots, N-1$: 
\begin{itemize}
\item[(i)]
$E^{(n)}$ ($n=1,2,\ldots,N$) are computed by solving Eq.~(\ref{recurrence_relation_triadsrg_D6}) iteratively with 
Eq.~(\ref{triad_env}). 
\item[(ii)]
Three procedures are repeated for $n=0,1,\ldots, N-1$:  
${\cal M}^{(n)}$ is decomposed as Eq.~(\ref{svd_M_triadsrg_D5})
using Eqs. (\ref{calU}), (\ref{calM}), (\ref{calD}) and (\ref{Me_triadSRG})
from $A^{(n)}, B^{(n)}, E^{(n+1)}$ and $U^{(n)}$. 
Then  $U^{(n)}$ is updated by 
Eq.~(\ref{X_triadsrg_D5}) and its SVD shown as Eqs. (\ref{x_svd}) and (\ref{u_update})
from $E^{(n+1)}, A^{(n)}, B^{(n)},U^{(n)}$. Then 
$A^{(n+1)}, B^{(n+1)}$ are updated by
Eqs. (\ref{new_triad_A}) and (\ref{new_triad_B}) from  $E^{(n+1)}, A^{(n)}, B^{(n)}$  and the new $U^{(n)}$.  
\end{itemize}
where $m$ is chosen so that converged or better results are obtained. 
We finally evaluate  
$Z \approx \sum_{i,j,a} A^{(N)}_{ija} B^{(N)}_{aji}$. 
The cost of this algorithm scales with ${\cal O} (m N\chi^5)$.

\section{Numerical tests \label{sec:numerical_results}}

We test the triad SRG method in the classical Ising model on a two dimensional square lattice
\footnote{
Let $\sigma_i$ be spin variable that takes $\sigma_i=\pm 1$ where $i$
 labels lattice sites of two dimensional square lattice. 
The Hamiltonian is given by 
$H=-J \sum_{\langle i,j\rangle} \sigma_i \sigma_j $ 
where  $\langle i,j\rangle$ are possible pairs of nearest neighbor sites. 
We can set $J=1$ without loss of generality. 
}.
The lattice volume is $V=2^{25} \times 2^{25}$ 
and the periodic boundary condition is 
assumed for two directions. 
The partition function,  which is defined in the standard manner as 
$Z = {\rm Tr} \, {\rm e}^{-H/T}$ with temperature $T$, is expressed as a tensor network 
Eq.(\ref{tensor_network}) with a $2 \times 2$ matrix $W$ given by
\begin{align}
    W &=
    \begin{pmatrix}
      \sqrt{\cosh(1/T)} & \sqrt{\sinh(1/T)}\\
      \sqrt{\cosh(1/T)} & -\sqrt{\sinh(1/T)}\\
    \end{pmatrix}.
\end{align}
The triad representation is introduced as Eq.(\ref{triad}) with identifications of Eq.~(\ref{AB_initial_triad})
where $W^{(i)}=W$ for $i=1,2,3,4$.

\begin{figure}[H]
\centering
\includegraphics[width=100mm]{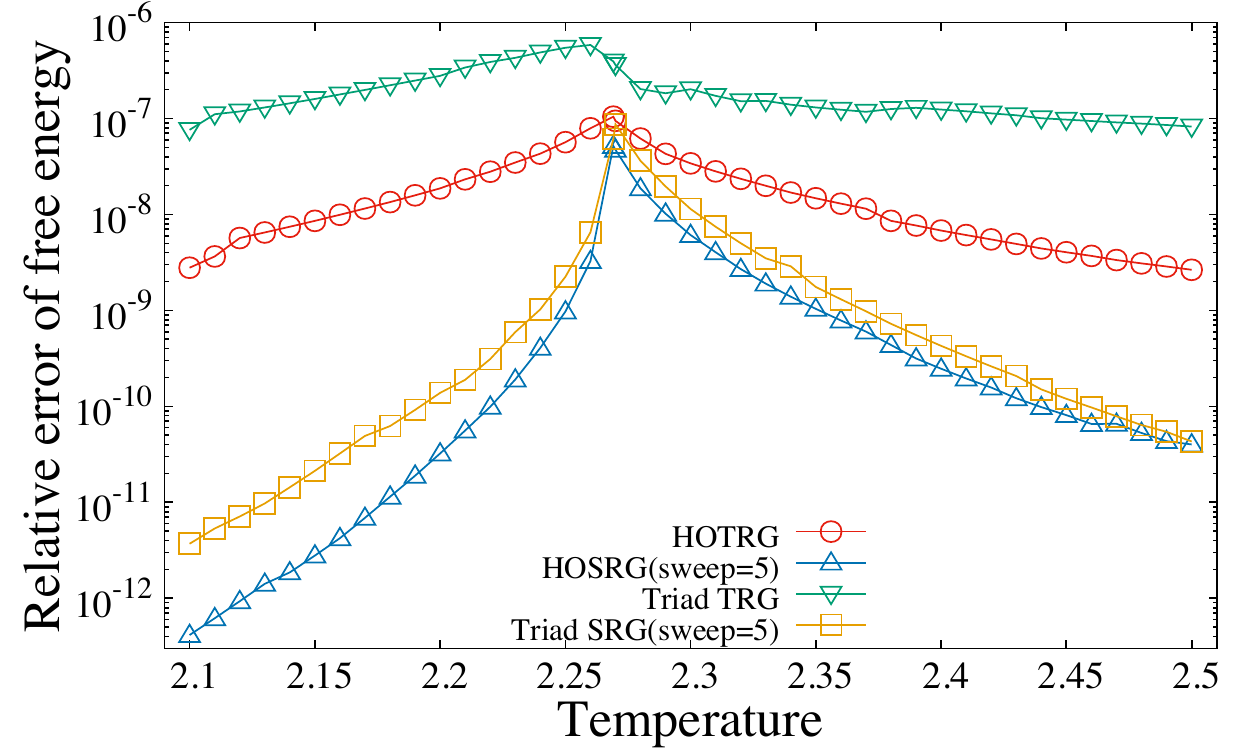}
\caption{
Relative error of the free energy against $T$ for $\chi=24$
}
 \label{fig:T_error_24}
\end{figure}

\begin{figure}[H]
  \centering
\includegraphics[width=100mm]{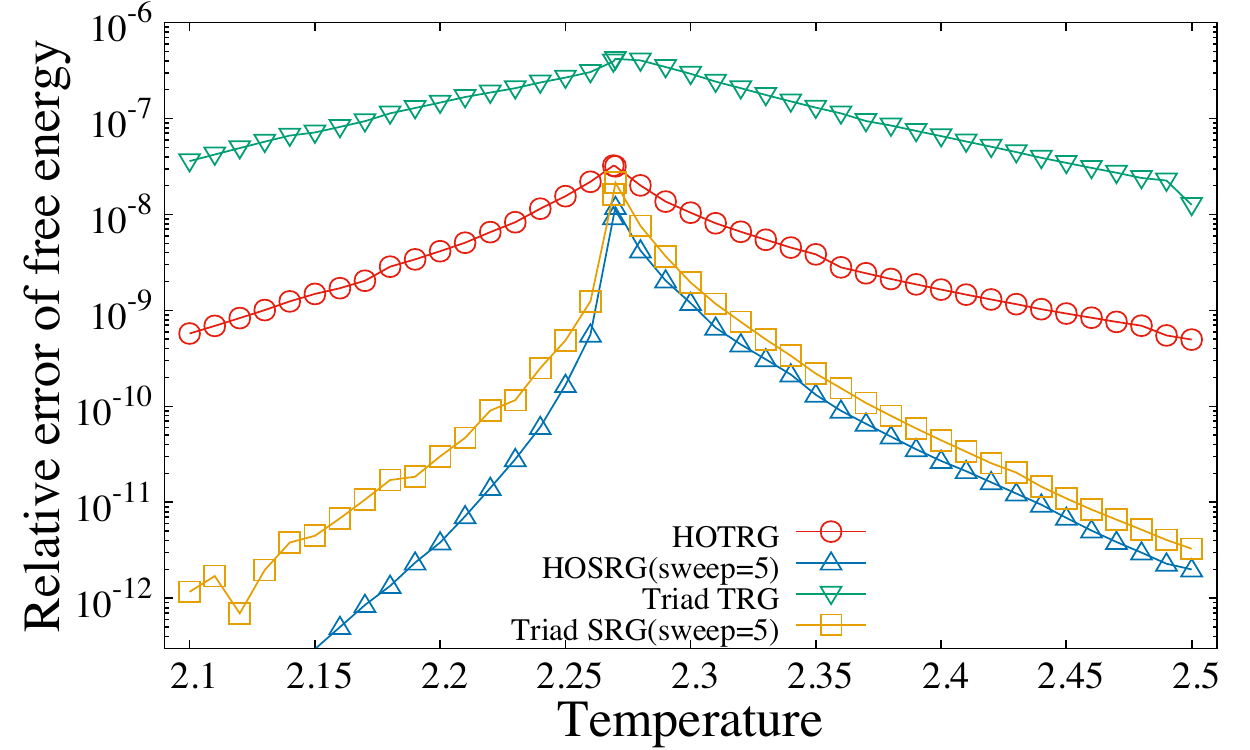}
\caption{
Relative error of the free energy against $T$ for $\chi=32$.
}
 \label{fig:T_error_32}
\end{figure}

\begin{figure}[H]
  \centering
\includegraphics[width=100mm]{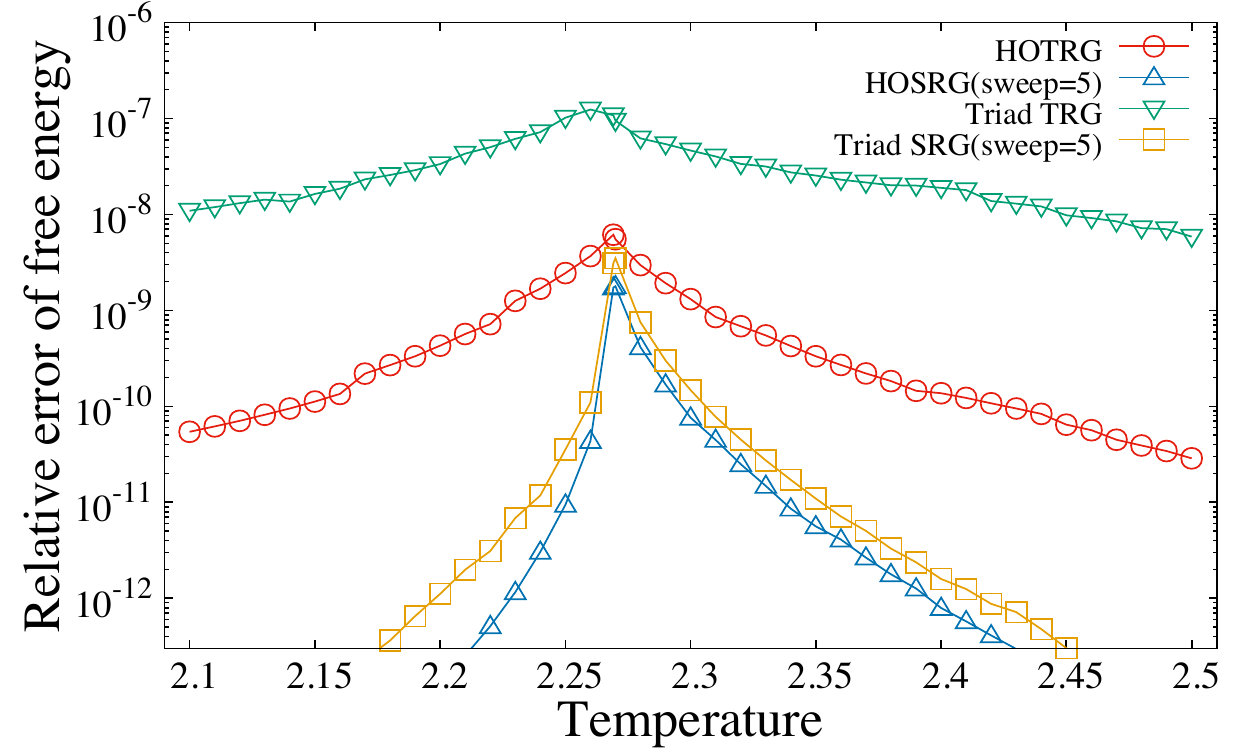}
\caption{
Relative error of the free energy against $T$ for $\chi=48$.
}
 \label{fig:T_error_48}
\end{figure}

The free energy is computed by the four different methods: HOTRG, HOSRG, triad TRG, and triad SRG. 
Let $\Delta F$ be the relative error  of free energy 
defined by
\begin{align}
\Delta F \equiv \left\vert \frac{F-F_{\rm trg}}{F} \right\vert 
\end{align}
where $F$ is the exact free energy and $F_{\rm trg}$ is the numerical result. 
Figs.~\ref{fig:T_error_24}--\ref{fig:T_error_48} show $\Delta F$ for $\chi=24,32,48$, respectively. 
In contrast to other methods with disentangler, like TNR etc,
SRG algorithms do not have any noticeable improvement at the critical point. 
See Fig.4 in ref.\cite{HOTRG}.
The error of the SRG schemes drastically reduces except near the critical point. 
For a fixed $\chi$, the triad SRG achieves better performance than the HOTRG and the triad TRG, 
but a little bit worse than the HOSRG. 

\begin{figure}[H]
  \begin{minipage}[b]{0.45\linewidth}
    \centering
    \includegraphics[width=70mm]{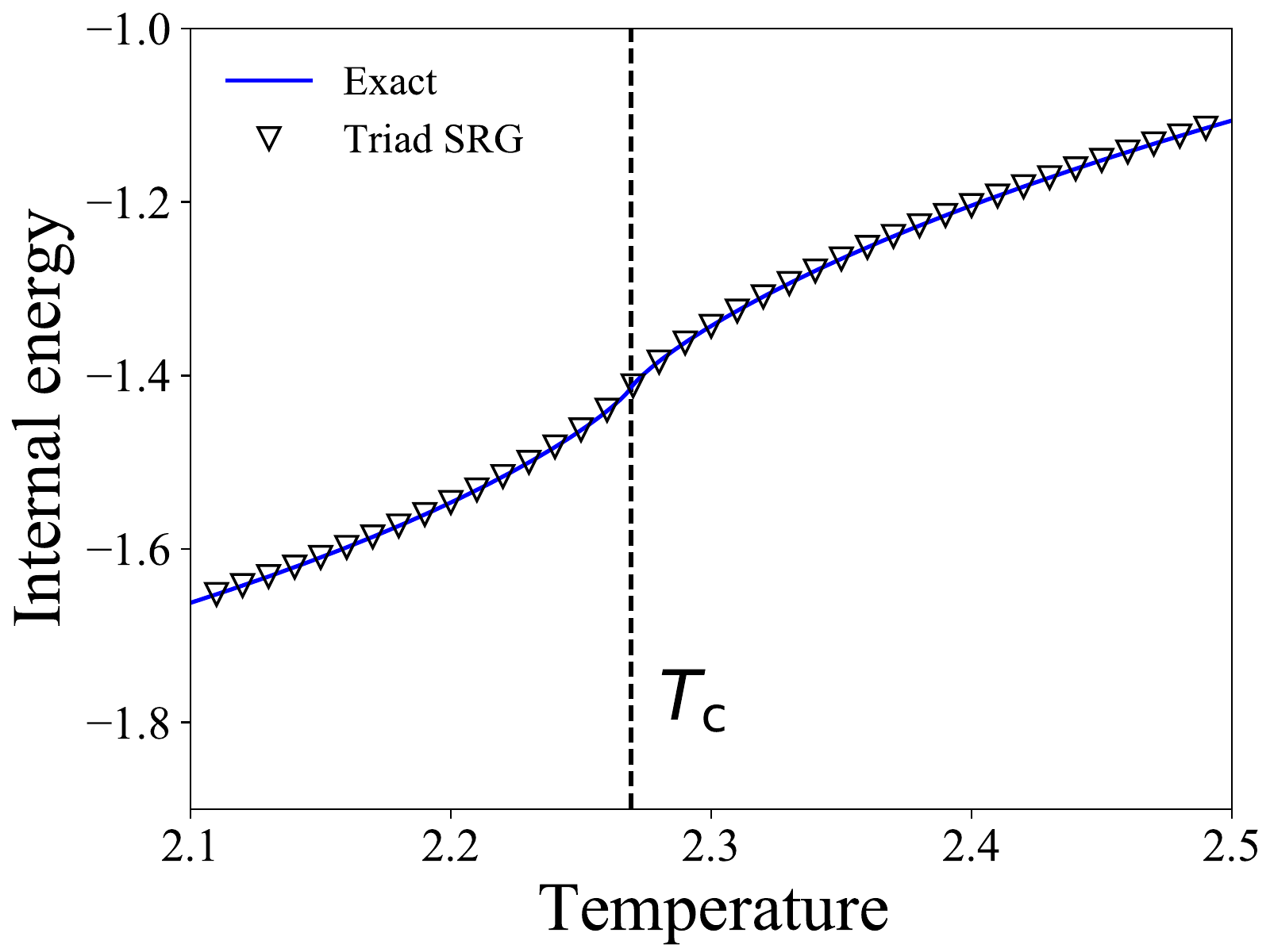}
    \subcaption{}
  \end{minipage}
  \begin{minipage}[b]{0.45\linewidth}
    \centering
    \includegraphics[width=70mm]{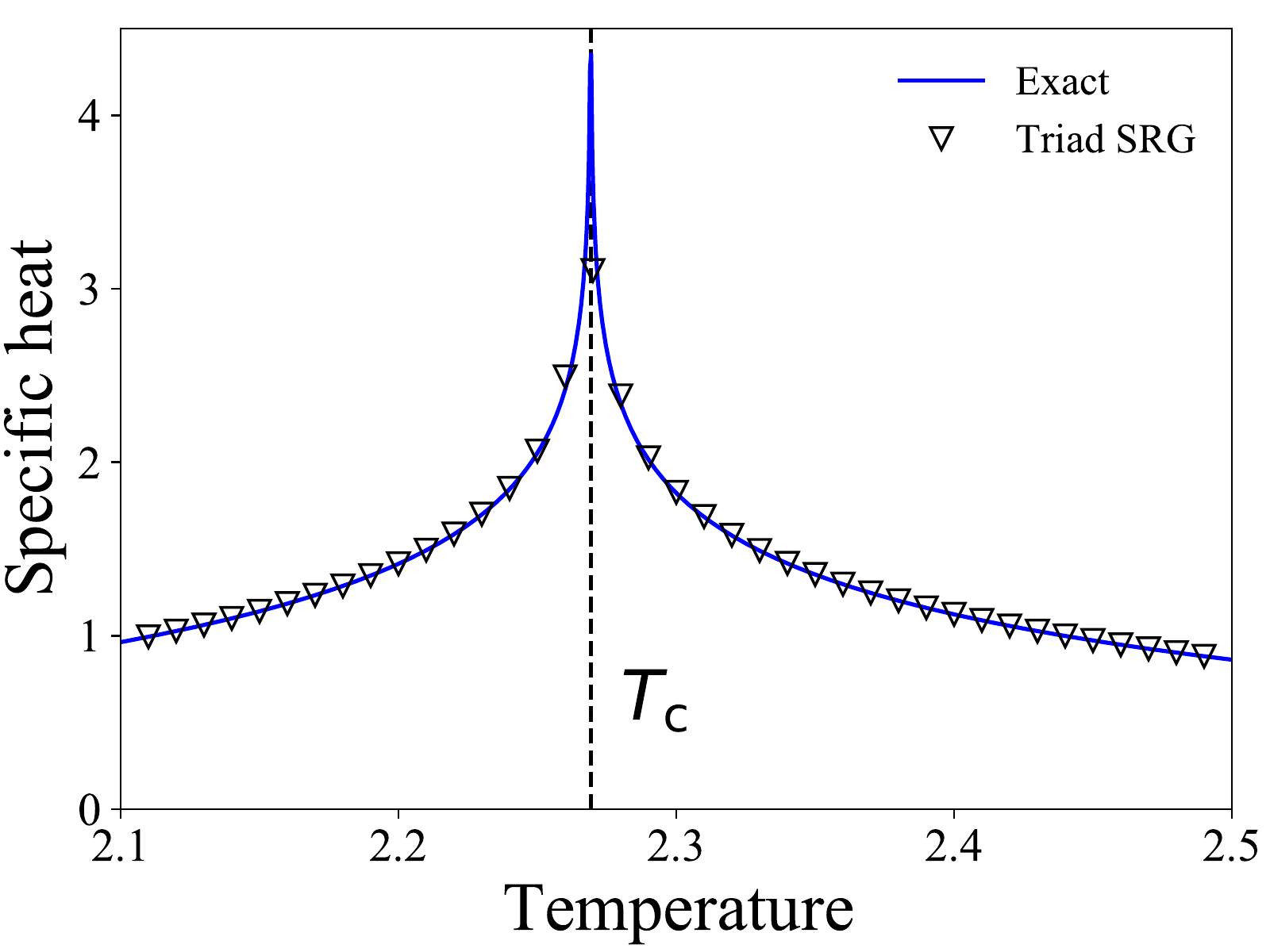}
    \subcaption{}
  \end{minipage}
  \caption{(a)The internal energy and (b)specific heat
    against $T$ for $\chi=48$.}
  \label{fig:internal_energy_specific_heat}
\end{figure}

Figs.~\ref{fig:internal_energy_specific_heat}~(a) and (b) show
the internal energy and specific heat obtained by the triad SRG method 
with the numerical derivatives from the free energy. 
We find that the results nicely reproduce the exact values. 

\begin{figure}[H]
\centering
\includegraphics[width=100mm]{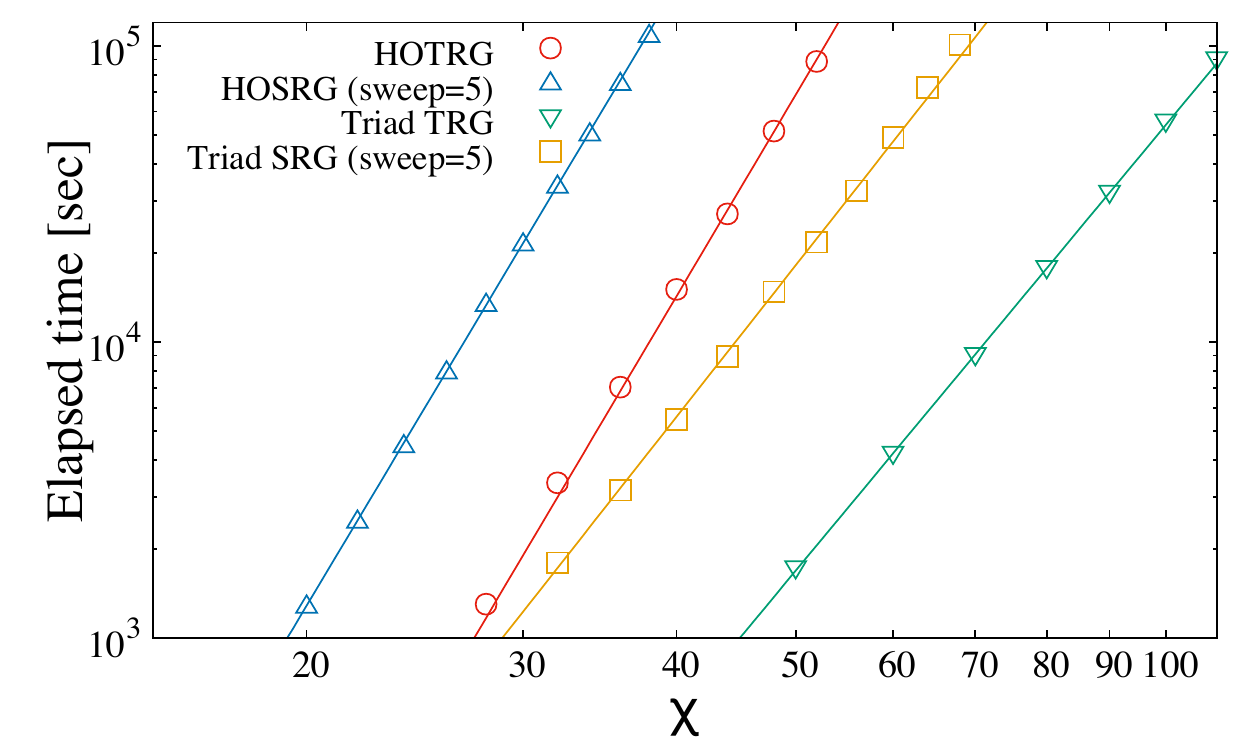}
\caption{
Elapsed time against the bond dimension. Solid lines are fit results using a power-law function $c \chi^p$.
The HOTRG and the HOSRG have $p \approx 7$ while
the triad TRG and triad SRG have $p \approx 5$.
}
 \label{fig:total_time_triad_srg}
\end{figure}

Fig.~\ref{fig:total_time_triad_srg} shows the $\chi$-dependence of the elapsed real time.  
Solid lines are  fit results using a power-law function $c \chi^p$. 
We find that $p \approx 7$ for the HOTRG/HOSRG and $p \approx 5$ 
for the triad TRG/triad SRG, and 
the theoretical $\chi$ dependence of the cost is nicely reproduced.  

\begin{figure}[H]
  \centering
  \includegraphics[width=100mm]{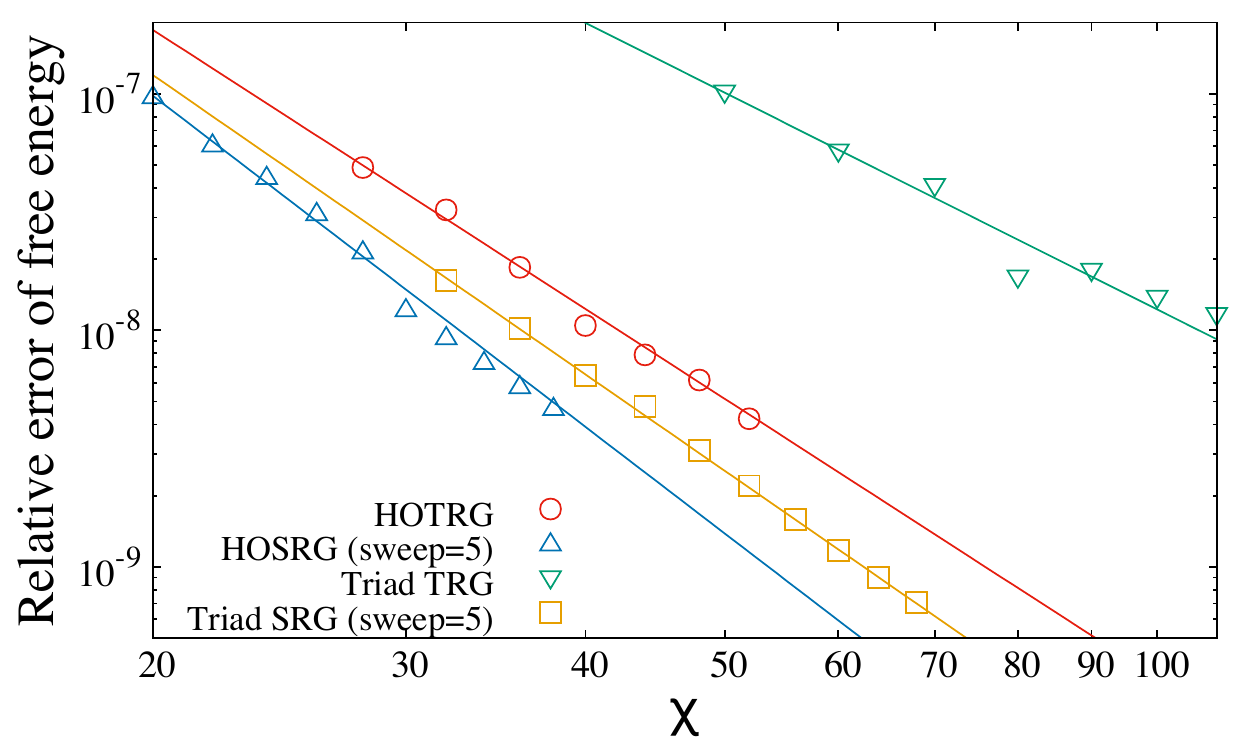}
\caption{
$\chi$ dependence of $\Delta F$ at $T=T_\cc$.
}
 \label{fig:dcut_error_tc}
\end{figure}

In Fig.~\ref{fig:dcut_error_tc}, we plot the relative error  of free energy against $\chi$ at $T_\cc$. 
The $\chi$ dependence of the errors are estimated
by power law fits as $error \simeq \chi^{-q}$. 
We obtain $ q \sim 4$ for the HOTRG, $q \sim 4.6$ for the HOSRG, $q\sim 3$ for the triad TRG 
and $q \sim 4.2$ for the triad SRG. 
Combining above results for Fig \ref{fig:total_time_triad_srg},  we have $error \sim (time)^{-q/p} $.
All the methods of Fig.~\ref{fig:dcut_error_tc} have $q/p>1/2$ better than the Monte Carlo methods  
that have $error \sim (time)^{-1/2}$ but worse than the TNR. 

\begin{figure}[H]
\centering
\includegraphics[width=100mm]{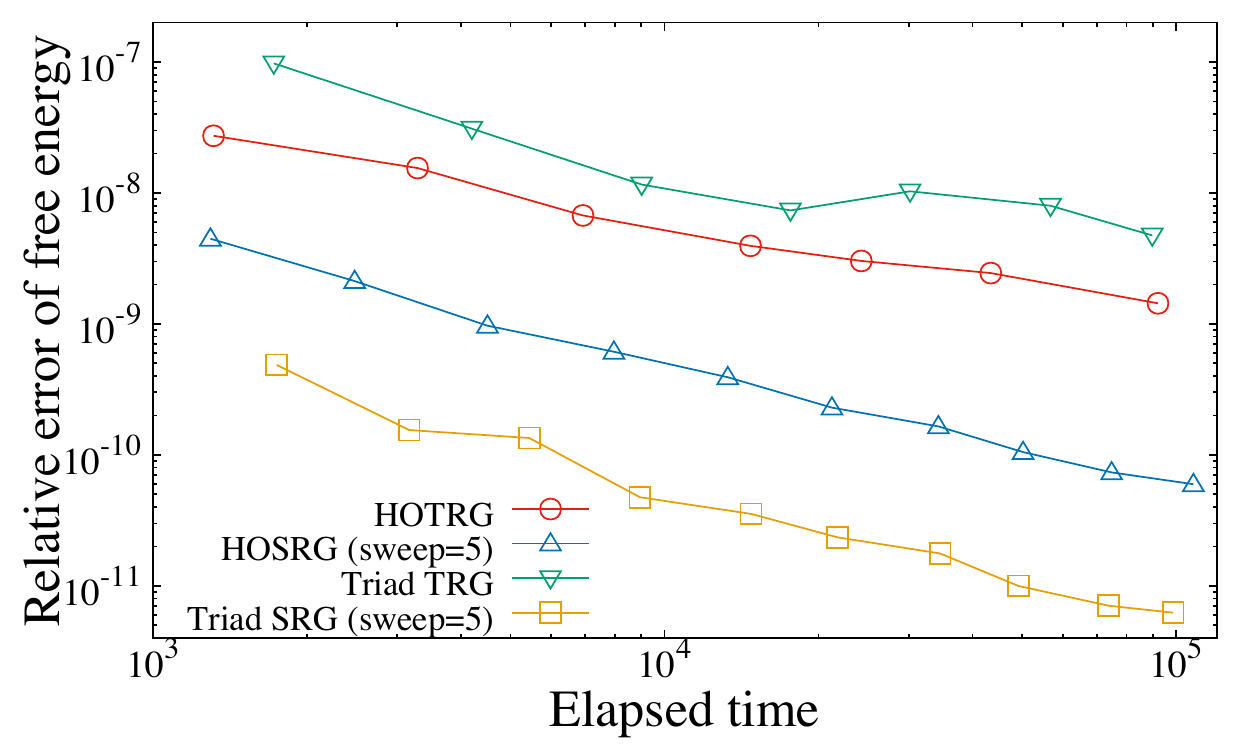}
\caption{
Time dependence of $\Delta F$ at $T=2.25 < T_\cc$.
}
 \label{fig:time_error_225}
\end{figure}

\begin{figure}[H]
  \centering
  \includegraphics[width=100mm]{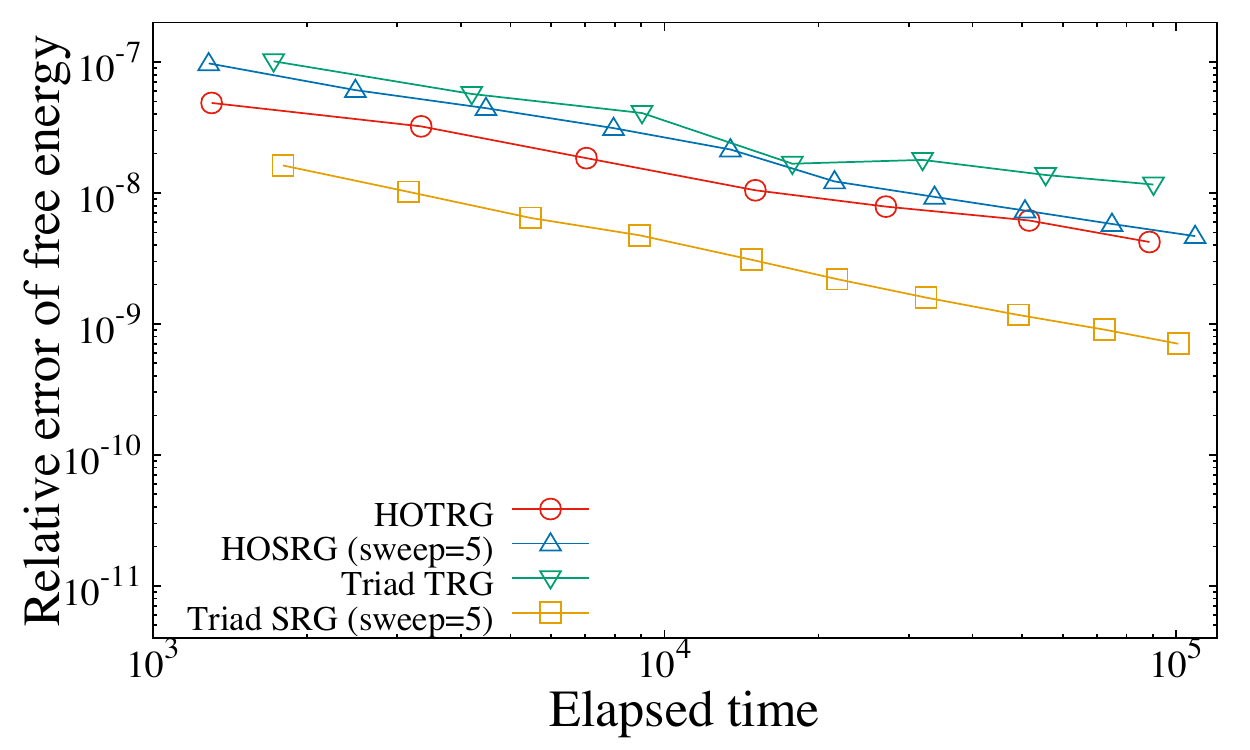}
\caption{
Time dependence of $\Delta F$ at $T=T_\cc$.
}
 \label{fig:time_error_tc}
\end{figure}

\begin{figure}[H]
  \centering
  \includegraphics[width=100mm]{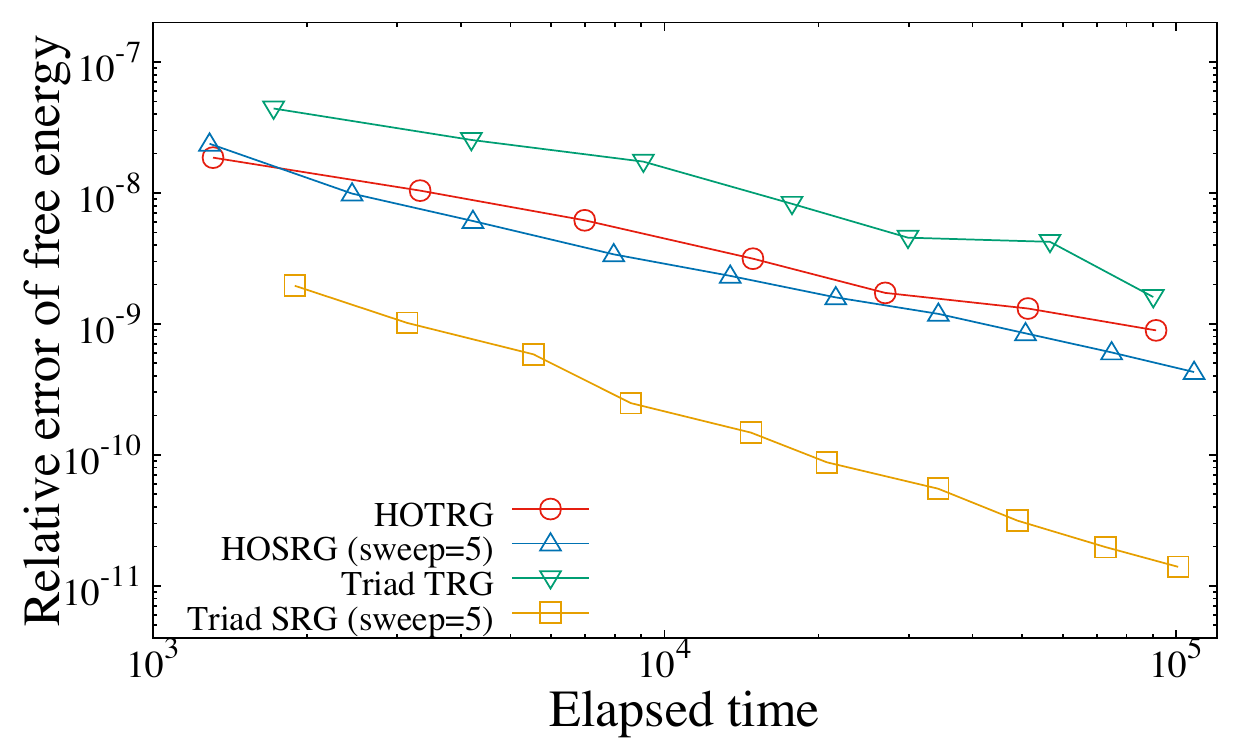}
\caption{
Time dependence of $\Delta F$ at $T=2.3 > T_\cc$. 
 }
 \label{fig:time_error_23}
\end{figure}

Figs.~\ref{fig:time_error_225}--\ref{fig:time_error_23} show 
the decrease of the error with the elapsed real time for $T=2.25, T_\cc, 2.3$, respectively. 
Since the triad method has low computational cost, 
the triad SRG has better performance than the other methods 
for the fixed execution time. 
We thus find that the SRG scheme and the triad representation of 
tensors coexist within sufficiently good accuracy. 

To obtain the data for
Figs.~\ref{fig:total_time_triad_srg}--\ref{fig:time_error_23}, 
we use a machine
which has 24GB for the memory 
and Intel(R) X5670 (2.93 Ghz 6 core) for the two CPUs.
The programs are written in python 2.7.5rc1,  
we use numpy.tensordot in numpy 1.8.0 for the tensor contractions and 
scipy.linalg.svd in scipy 0.14.0 for the SVD of the tensors.

\section{Summary and discussion\label{sec:discussion}}

We presented a second renormalization group method 
in a two dimensional triad network called the triad SRG. 
The $\chi^6$-algorithm was given 
by applying the SRG to two parts of the triad TRG, which are 
the decompositions of intermediate tensors  
and the preparation of isometries. 
Since the environment tensors are given by rank-3 tensors, 
the $\chi^5$-algorithm can also be given by using the randomized SVD. 

The numerical results show that the triad SRG has better performance 
than the HOTRG/HOSRG and the triad TRG for the fixed computational time. 
Although the results do not show better performance than the TNR, 
we find that the SRG-scheme  and the triad representation of the environment tensors 
coexist within good accuracy.

In Ref.\cite{Morita2020}, the influence of environments is incorporated into 
the calculations without any backward scheme. With such a technique, 
it could be possible to further reduce the cost of the triad SRG. 
Since the  triad representation of $2d$-rank tensor is not unique, 
we have to find the better representation of tensors and environments  
to apply this method to higher dimensions. 
Then techniques mentioned in this paper will help us to improve higher dimensional algorithms.

Our paper is devoted for making the SRG method with the triad representations of tensor.  
This is the first step to extend our method in higher dimensions. 
It is not straightforward to evaluate a computational cost in $d$-dimensions rigorously 
because the triad representation is not unique in higher dimensions. However, 
we roughly estimate it as $O(N\chi^{d+3})$, where $N$ is the number of forward/backward steps and $\chi$ is the bond dimension.
This is derived from a naive extension from $2d$ triad SRG, whose cost is $O(N\chi^5)$, to $d$-dimensions.
The results of this paper and the cost estimation could help us to make a better TRG method in the future.

\begin{acknowledgments}
This work was supported by JSPS KAKENHI Grant Numbers 17K05411, 19K03853 and 21K03531. 
D.K. would like to thank David C.-J. Lin and the members of NCTS in National Tsing-Hua University 
for encouraging me.

\end{acknowledgments}

\appendix

%\bibliography{ref}

%\addcontentsline{toc}{chapter}{\bibname}

\end{document}